\documentclass[11pt,a4paper]{article}

\usepackage{graphicx,latexsym,color,amsfonts}
\usepackage{cite,citesort}
\usepackage{amsmath}

\newif\ifgreek
\def\testgreek#1{
  \ifx#1\gamma\greektrue\else\ifx#1\Gamma\greektrue\else
  \ifx#1\epsilon\greektrue\else
  \ifx#1\mu\greektrue\else
  \ifx#1\rho\greektrue\else
  \ifx#1\sigma\greektrue\else\ifx#1\Sigma\greektrue\else
     \greekfalse
  \fi\fi\fi\fi\fi\fi\fi}

\newcommand{\mat}[1]{{\testgreek#1\ifgreek\boldsymbol#1\else
                      \mathbf#1\fi}} 

\newcommand{\R}{\mathbb{R}}

\newcommand{\ie}{\textit{i.e.}\/, }
\newcommand{\eg}{\textit{e.g.}\/, }
\newcommand{\cf}{\textit{cf.}\/, }

\providecommand*{\mrm}[1]{\mathrm{#1}}

\providecommand*{\unit}[1]{\ensuremath{\mrm{\,#1}}}
\providecommand*{\eu}{\ensuremath{\mrm{e}}}
\providecommand*{\iu}{\ensuremath{\mrm{i}}}
\renewcommand{\Re}{\operatorname{Re}}	
\renewcommand{\Im}{\operatorname{Im}}	
\providecommand*{\diff}{\operatorname{d}\!}

\renewcommand{\vec}[1]{{\boldsymbol#1}}
\newcommand{\myvec}[1]{{\boldsymbol#1}}
\newcommand{\ju}{\mrm{j}}
\newcommand{\clight}{\mrm{c}_0}

\newcommand{\partder}[2]{\frac{\partial#1}{\partial#2}}

\newcommand{\Ev}{\vec{E}}

\newcommand{\Fv}{\vec{F}}
\newcommand{\Bv}{\vec{B}}
\newcommand{\Jv}{\vec{J}}

\newcommand{\Av}{\vec{A}}

\newcommand{\Hv}{\vec{H}}

\newcommand{\rv}{\vec{r}}
\newcommand{\qv}{\vec{q}}

\newcommand{\dv}{\vec{d}}

\newcommand{\Wv}{\vec{W}}

\newcommand{\rvh}{\hat{\vec{r}}}
\newcommand{\nvh}{\hat{\vec{n}}}

\newcommand{\xvh}{\hat{\vec{x}}}
\newcommand{\yvh}{\hat{\vec{y}}}
\newcommand{\zvh}{\hat{\vec{z}}}

\newcommand{\dtau}{\bar{\tau}}

\newcommand{\Lop}{\mathop{\mathcal{L}}\nolimits}

\newcommand{\Yop}{\mathop{\mrm{Y}}\nolimits}
\newcommand{\Yvop}{\mathop{\mat{Y}}\nolimits}
\newcommand{\jop}{\mathop{\mrm{j}}\nolimits}

\newcommand{\nop}{\mathop{\mrm{n}}\nolimits}
\newcommand{\hop}{\mathop{\mrm{h}}\nolimits}

\newcommand{\uop}{\mathop{\mat{u}}\nolimits}

\newcommand{\Rop}{\mathop{\mrm{R}}\nolimits}

\newcommand{\Id}{\mat{I}}
\newcommand{\Gm}{\mat{G}}

\newcommand{\qtext}[1]{\quad\text{#1}\ } 

\newcommand{\diffS}{\mathop{\mathrm{\mathstrut{dS}}}\nolimits}
\newcommand{\diffV}{\mathop{\mathrm{\mathstrut{dV}}}\nolimits}
\newcommand{\diffVa}{\diffV_{\!1}}
\newcommand{\diffVb}{\diffV_{\!2}}

\newcommand{\refl}{\varGamma}

\newcommand{\We}{{W}^{\mrm{(E)}}}
\newcommand{\Wm}{{W}^{\mrm{(M)}}}

\newcommand{\Ordo}{\mathcal{O}}

\newcommand{\Rr}{\R_{\mrm{r}}}

\newcommand{\WE}{W^{\mrm{(E)}}}
\newcommand{\WM}{W^{\mrm{(M)}}}

\newcommand{\WFE}{W_{\mrm{F}}^{\mrm{(E)}}}
\newcommand{\WFM}{W_{\mrm{F}}^{\mrm{(M)}}}
\newcommand{\WPE}{W_{\mrm{P}}^{\mrm{(E)}}}
\newcommand{\QE}{Q^{\mrm{(E)}}}
\newcommand{\QM}{Q^{\mrm{(M)}}}

\newcommand{\QZp}{Q_{\mrm{Z'}}}
\newcommand{\QF}{Q_{\mrm{F}}}
\newcommand{\QP}{Q_{\mrm{P}}}
\newcommand{\QB}{Q_{\mrm{B}}}
\newcommand{\Qa}{\langle Q\rangle}
\newcommand{\QZpE}{Q_{\mrm{Z'}}^{\mrm{(E)}}}
\newcommand{\QFE}{Q_{\mrm{F}}^{\mrm{(E)}}}
\newcommand{\QBE}{Q_{\mrm{B}}^{\mrm{(E)}}}
\newcommand{\QZpM}{Q_{\mrm{Z'}}^{\mrm{(M)}}}
\newcommand{\QFM}{Q_{\mrm{F}}^{\mrm{(M)}}}
\newcommand{\QBM}{Q_{\mrm{B}}^{\mrm{(M)}}}

\newcommand{\dB}{\unit{dB}}

\title{Stored Electromagnetic Energy and Antenna Q}

\title{Stored Electromagnetic Energy and Antenna Q}
\author{Mats Gustafsson\thanks{Department of Electrical and Information Technology, Lund University, Box 118, SE-221 00 Lund, Sweden. (Email: mats.gustafsson@eit.lth.se).}\ \ and B. L. G. Jonsson\thanks{KTH Royal Institute of Technology, School of Electrical Engineering, Department of Electromagnetic Engineering, Teknikringen 33, SE-100 44 Stockholm, Sweden. (Email: lars.jonsson@ee.kth.se)}}

\begin{document}

\maketitle

\begin{abstract}
	Decomposition of the electromagnetic energy into its stored and radiated parts is instrumental in the evaluation of antenna Q and the corresponding fundamental limitations on antennas. This decomposition is not unique and there are several proposals in the literature.
	Here, it is shown that stored energy defined from the difference between the energy density and the far field energy equals the new energy expressions proposed by Vandenbosch for many cases. This also explains the observed cases with negative stored energy and suggests a possible remedy to them. The results are compared with the classical explicit expressions for spherical regions where the results only differ by ka that is interpreted as the far-field energy in the interior of the sphere. Numerical results of the Q-factors for dipole, loop, and inverted L-antennas are also compared with estimates from circuit models and differentiation of the impedance. The results indicate that the stored energy in the field agrees with the stored energy in the Brune synthesized circuit models whereas the differentiated impedance gives a lower value for some cases. The corresponding results for the bandwidth suggest that the inverse proportionality between bandwidth and Q depends on the relative bandwidth or equivalent the threshold of the reflection coefficient. The Q from the differentiated impedance and stored energy are most useful for relative narrow and wide bandwidths, respectively.
\end{abstract}

\section{Introduction}
Electrostatic energy in free space can be written as an integral of the energy density, $\epsilon_0|\Ev|^2/4$, or equivalently as an integral of the electric potential, $\phi$, times the charge density, $\rho$,~\cite{Jackson1999,Landau+Lifshitz1984,Feynman1965,vanBladel2007}. A similar expression holds for the magnetostatic energy. The electrodynamic case is more involved. In~\cite{Carpenter1989}, Carpenter suggests a generalization in the time domain based on $\phi\rho+\Av\cdot\Jv$, \ie the sum of the scalar potential times the charge density and the vector potential, $\Av$, times the electric current density, $\Jv$, see also~\cite{Endean+Carpenter1992,Uehara+etal1992}.  Geyi uses a similar expression to analyze small antennas in~\cite{Geyi2003}.
	Vandenbosch presents general integral expressions in the electric current density for the stored electric and magnetic energies in the frequency domain~\cite{Vandenbosch2010} and time domain~\cite{Vandenbosch2013a,Vandenbosch2013b}. These expressions are similar to the expressions by Carpenter but include some correction terms. 
	
Stored electromagnetic energy is instrumental in determination of lower bounds on the Q-factor for antennas, see~\cite{Volakis+etal2010} for an overview. The classical results by Chu~\cite{Chu1948} and Collin \& Rothschild~\cite{Collin+Rothschild1964} are based on subtraction of the power flow and explicit calculations using mode expansions of the stored energy outside a sphere. This gives simple expressions for the minimum $Q$ of small spherical antennas~\cite{Chu1948,Collin+Rothschild1964}. The major shortcoming is that the results are restricted to spherical regions although some generalizations to spheroidal regions are suggested in~\cite{Foltz+McLean1999,Sten+etal2001}. The results have also been generalized to the case with electric current sheets by Thal~\cite{Thal2006} and Hansen and Collin~\cite{Hansen+Collin2009} by adding the stored energy in the interior of the sphere. Yaghjian and Best~\cite{Yaghjian+Best2005} analyze stored energy for general media and its relation to the input impedance. The new energy expressions by Vandenbosch~\cite{Vandenbosch2010} are very useful as they express the stored energy in the current density on the antenna structure. This is very fruitful  
in the analysis of small antennas~\cite{Vandenbosch2011,Gustafsson+etal2012a,Capek+etal2012,Gustafsson+Nordebo2013} and for antenna optimization~\cite{Gustafsson+Nordebo2013,TEAT-7227}. 
The expressions have been verified for wire antennas in~\cite{Hazdra+etal2011}. One minor problem with the proposed expressions is that they can produce negative values of stored energy for electrically large structures~\cite{Gustafsson+etal2012a}. A similar relation with the differentiated antenna input impedance~\cite{Yaghjian+Best2005,Gustafsson+Nordebo2006b} is derived in~\cite{Capek+etal2014}.

In this paper, we investigate stored electric and magnetic energy expressions based on subtraction of the far-field energy density. The expressions are suitable for antenna Q and bandwidth calculations and they are closely related to the classical methods in~\cite{Collin+Rothschild1964,Yaghjian+Best2005}, see also~\cite{Volakis+etal2010}, for antenna Q calculations. They are not restricted to spherical geometries and, furthermore, resembles the recently proposed expressions by Vandenbosch~\cite{Vandenbosch2010}. The results provide a new interpretation of Vandenbosch's expressions~\cite{Vandenbosch2010} and explain the observed cases with negative stored energy~\cite{Gustafsson+etal2012a}. They also suggest a possible remedy to the negative energy and that the computed $Q$ has an uncertainty of the order $ka$, where $a$ is the radius of the smallest circumscribing sphere and $k$ the wavenumber. This is consistent with the use of $Q$ for small (sub wavelength) antennas, where $ka$ is small and $Q$ is large~\cite{Chu1948,Collin+Rothschild1964}.
Analytic results for spherical structures show that the expressions in~\cite{Vandenbosch2010} for $Q$ differ with $ka$ from the results in~\cite{Hansen+Collin2009}, that is here interpreted as the far-field energy in the interior of the sphere. The results for $Q$ are also compared with estimated values from circuit models and differentiation of the impedance~\cite{Yaghjian+Best2005,Gustafsson+Nordebo2006b} for dipole, loop, and inverted L antennas.

We use Brune synthesis~\cite{Wing2008} to construct equivalent lumped circuit models from the input impedance. The numerical results indicate that the stored energy in the circuit elements agree very well with the stored energy in the fields. The results also show that the Q-factor from differentiation of the input impedance, $\QZp$, agree with the $Q$ from the stored energy, $\QB$, if $Q$ is large and dominated by a single resonance. The values start to differ for lower values of $Q$ where multiple resonances are common~\cite{Gustafsson+Nordebo2006b,Stuart+etal2007}. We also compare the corresponding bandwidth with and without matching networks. The results indicate that the inverse proportionality between the fractional bandwidth $B\sim 1/Q$ is most accurate using $Q=\QZp$ for relative narrow bandwidths $B<2/Q$ whereas $Q=\QB$ is better for wider bandwidths. This is consistent with $\QZp$ being a local function of the input impedance and $\QB$ depending on the global behavior of the input impedance. The bandwidth for a simple shunt and series resonance circuit~\cite{Gustafsson+Nordebo2006b} is also analyzed using matching networks and Fano matching bounds~\cite{Fano1950} to illustrate a case with $\QZp=0$, where the inverse proportionality of the bandwidth to Q fails for $\QZp$.

The paper is organized as follows. In Sec.~\ref{S:EMenergy}, the stored electric and magnetic energies defined by subtraction of far-field from the energy density are analyzed. The coordinate dependence is analyzed in Sec.~\ref{S:coordinatedependent}. Stored energies from small structures are derived in Sec.~\ref{S:small}. Stored energy from the input impedance are discussed in Sec.~\ref{S:Zenergy}.
Analytic results for spherical geometries and resonance circuits and numerical results for dipole, loop, and inverted L antennas are presented in Sec.~\ref{S:antennaex}. The paper is concluded in Sec.~\ref{S:conclusions}. There are four appendices discussing Green's function identities App.~\ref{sec:GFI}, spherical geometries App.~\ref{S:Sphere}, Brune synthesis App.~\ref{S:Brune}, and Bode-Fano matching App.~\ref{S:BodeFano}.

\section{Stored electromagnetic energy}\label{S:EMenergy}
We consider time-harmonic electric and magnetic fields, $\Ev(\rv)$ and $\Hv(\rv)$, respectively, with a suppressed $\eu^{-\iu\omega t}$ dependence, where $\omega$ denotes the angular frequency. The Maxwell equations in free space are \cite{Jackson1999}
\begin{equation}\label{eq:maxwelleqs}
\begin{cases}
	\nabla\times\Ev = \iu\omega\mu_0\Hv = \iu\eta_0 k\Hv\\
	\nabla\times\Hv = -\iu\omega\epsilon_0\Ev +\Jv = -\frac{\iu k}{\eta_0}\Ev + \Jv, 
\end{cases}
\end{equation}
where $\Jv$ denotes the current density, while $\epsilon_0$, $\mu_0$, and $\eta_0=\sqrt{\mu_0/\epsilon_0}$ are the free space permittivity, permeability, and impedance, respectively. 
For simplicity, we interchange between the angular frequency and the free space wavenumber $k=\omega/\clight$, where the speed of light  $\clight=1/\sqrt{\mu_0\epsilon_0}$. We also recall the continuity equation, $\nabla\cdot\Jv=\iu\omega\rho$, relating the current density $\Jv$ with the charge density $\rho$.

The time-harmonic electric and magnetic energy densities~\cite{Jackson1999,Landau+Lifshitz1984,vanBladel2007} are $\epsilon_0|\Ev|^2/4$ and $\mu_0|\Hv|^2/4$, respectively. The energy densities are not observable~\cite{Carpenter1989} and there are a few alternative suggestions in the literature~\cite{Feynman1965}. The electric and magnetic energies comprise both radiated and stored energies; however, for antenna Q calculations one must extract the stored energy. 
\begin{figure}[t]
  \begin{center}
  \includegraphics[width=0.35\textwidth]{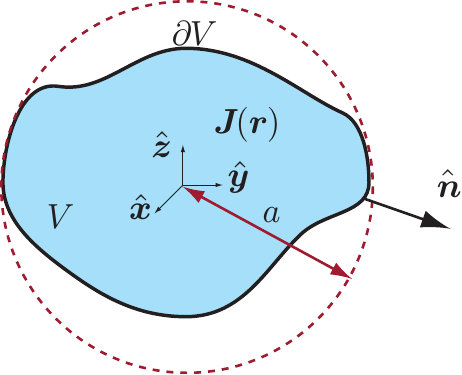}
  \end{center}
  \caption{Illustration of the object geometry $V$ with surface $\partial V$, outward normal unit vector $\nvh$, and current density $\Jv(\rv)$. The object is circumscribed by a sphere with radius $a$.}
  \label{fig:objectgeo}
\end{figure}

The Maxwell's equations~\eqref{eq:maxwelleqs} show that the sources and fields obey the conservation of energy equation in differential form, 
\begin{equation}\label{eq:conserv_energ1}
	\iu 2\omega \big(\frac{\epsilon_0}{4}|\Ev|^2-\frac{\mu_0}{4}|\Hv|^2\big)+\frac{1}{2} 	\Ev\cdot\Jv^{\ast}
	=-\frac{1}{2} \nabla\cdot(\Ev\times\Hv^{\ast}), 
\end{equation}
where the superscript ${}^\ast$ denotes complex conjugate. We consider current distributions $\Jv$ whose support is in a volume $V$ bounded by the surface $\partial V$, see Fig.~\ref{fig:objectgeo}. Integrating \eqref{eq:conserv_energ1} over this volume gives the real part result 
\begin{equation}\label{eq:energyid_Re}
	 \Re\int_{\partial V} \frac{\Ev(\rv)\times\Hv^{\ast}(\rv)\cdot\nvh(\rv)}{2}\diffS
	= -\Re\int_V\frac{\Ev(\rv)\cdot\Jv^{\ast}(\rv)}{2}\diffV,
\end{equation}
where $\nvh$ denotes the outward-normal unit vector of the surface $\partial V$. The first term in the real part expression~\eqref{eq:energyid_Re} is readily identified in view of the Poynting vector~\cite{Jackson1999,vanBladel2007} as the time-average radiated power flow through the surface $\partial V$, so that \eqref{eq:energyid_Re} equates the radiated power exiting $\partial V$ to the time average of the power generated by $\Jv$, as expected from energy conservation.  
Furthermore, integrating~\eqref{eq:conserv_energ1} over all space shows that the radiated power exiting the surface $\partial V$ can be expressed in terms of the far field as
\begin{equation}\label{eq:radiatedpower}
	P_{\mrm{r}}=\Re\int_{\partial V} \frac{\Ev(\rv)\times\Hv^{\ast}(\rv)\cdot\nvh(\rv)}{2}\diffS
	=\int_{\Omega}\frac{|\Fv(\rvh)|^2}{2\eta_0}\diff\Omega,
\end{equation}
where $\Omega$ denotes the surface of the unit sphere and the far field behaves like $\Ev(\rv)\sim\eu^{\iu kr}\Fv(\rvh)/r$ as $r\to\infty$, where $\rv=r\rvh$ and $r=|\rv|$. Similarly, by integrating~\eqref{eq:conserv_energ1} over all space one obtains 
the imaginary part result
\begin{equation}\label{eq:energyid_Im}
	\int_{\R^3}\!\frac{\mu_0}{4}|\Hv(\rv)|^2-\frac{\epsilon_0}{4}|\Ev(\rv)|^2 \diffV
	=\Im\int_V\!\frac{\Ev(\rv)\cdot\Jv^{\ast}(\rv)}{4\omega}\diffV,
\end{equation}
where we used the fact that the integral of the imaginary part of the divergence term in~\eqref{eq:conserv_energ1} vanishes as the integration volume approaches $\R^3$. The imaginary part result~\eqref{eq:energyid_Im} relates the well-defined difference between the time-average electric and magnetic energies with the net reactive power delivered by $\Jv$.

As is well known \cite{Chu1948,Collin+Rothschild1964}, the total energy, defined as the integral of the energy density integrated over all space, is unbounded due to the $1/r^2$ decay of the energy density in the far radiation zone. This is resolved by decomposition of the total energy into radiated and stored energy. The stored energy is, however, difficult to define and interpret. The classical approach used by Chu~\cite{Chu1948} and Collin \& Rothschild~\cite{Collin+Rothschild1964}, and subsequently by others, is based on mode expansions, and therefore restricted to canonical geometries. Spherical regions are most commonly considered but there are also some results for cylindrical~\cite{Collin+Rothschild1964} and spheroidal~\cite{Foltz+McLean1999,Sten+etal2001} structures. The stored energy density is customarily defined as the difference between the total energy density and the \textit{radiated power flow} in the radial direction~\cite{Collin+Rothschild1964,McLean1996,Collin1998}, thus the stored electric energy becomes
\begin{equation}\label{eq:storedEP}
		\WPE
	= \frac{\epsilon_0}{4}\int_{\Rr^3}\! |\Ev(\rv)|^2
	-\eta_0\Re\{\Ev(\rv)\times\Hv^{\ast}(\rv)\}\cdot\rvh\diffV,
\end{equation}
where $\Rr^3=\{\rv: \lim_{r_0\to\infty}|\rv|\leq r_0\}$ is used to indicate that the integration is over an infinite spherical volume.
The classical results by Chu~\cite{Chu1948} are for spheres with vanishing interior field~\cite{Collin+Rothschild1964},  so that the 
stored energy is due to the exterior field only (\ie for the region where $r>a$ where $a$ is the radius of the smallest sphere circumscribing the sources). The Thal bound~\cite{Thal2006} generalizes the results to fields generated by electric surface currents, see also~\cite{Hansen+Collin2009}. Here it is observed that there is a stored energy but no radiated energy flux in the interior of the sphere. The definition~\eqref{eq:storedEP} is useful for spherical geometries and can be generalized to cylindrical geometries~\cite{Collin+Rothschild1964,McLean1996,Collin1998}. This definition is difficult to generalize to arbitrary geometries due to its coordinate dependence that originates from the scalar multiplication with $\rvh$. The subtraction of the radiated energy flow is equivalent to subtraction of the energy of the far field outside a circumscribing sphere, \cf \eqref{eq:radiatedpower}. This suggests an alternative stored electric energy defined by subtraction of the \textit{far-field energy}, \ie
\begin{equation}\label{eq:storedEF}
	\WFE
	=\frac{\epsilon_0}{4}\int_{\Rr^3} |\Ev(\rv)|^2-\frac{|\Fv(\rvh)|^2}{r^2} \diffV,
\end{equation}
where the integration is over the infinite sphere $\Rr^3$. The subtracted far-field in the integrand can alternatively be written as a subtraction of the radius times the radiated power~\cite{Yaghjian+Best2005}. 

We note that the definitions with the power flow~\eqref{eq:storedEP} and far field~\eqref{eq:storedEF} differ only in the interior of the smallest circumscribing sphere associated with the source support. In the interior of the smallest circumscribing sphere, which we assume next to be of radius $a$, this subtracted far-field energy is then
\begin{equation}
	\frac{\epsilon_0}{4}\int_{0}^a\int_\Omega|\Fv(\rvh)|^2\diff\Omega\diff r
	=\frac{a}{2\clight}P_{\mrm{r}}.
\end{equation}
Assuming that the contribution to the true stored electric energy, say $\WE$, due to the exterior field outside the smallest circumscribing sphere, is equal to that of $\WPE$ and $\WFE$ in~\eqref{eq:storedEP} and~\eqref{eq:storedEF}, and that it subtracts some non-negative value less than $\epsilon_0|\Fv|^2/(4r^2)$ inside the sphere, then we obtain the bound
\begin{equation}\label{eq:energybound}
	\WFE \leq \WE  \leq \WFE
	+\frac{a}{2\clight}P_{\mrm{r}}.
\end{equation}
This means that the stored electric energy can be bounded from below and above by~\eqref{eq:storedEF}.
The stored magnetic energy, $\WFM$, is defined analogously. The stored energy is commonly normalized with the radiated power to define Q-factors.
The Q-factor is $Q = \max\{\QE,\QM\}$, where
\begin{equation}\label{eq:QEQM}
	\QE=\frac{2\omega \WE}{P_{\mrm{r}}}
	\qtext{and }
	\QM=\frac{2\omega \WM}{P_{\mrm{r}}}
\end{equation}
and we have included a factor of 2 in the definitions of $\QE$ and $\QM$  to simplify the comparison with antenna $Q$.
This translates the bound~\eqref{eq:energybound} into
\begin{equation}\label{eq:Qbound}
	\max\{0,\QF\} \leq Q \leq \QF + ka,
\end{equation}
where we have added that $Q$ is non-negative.

We show that the stored energy with the subtracted far field~\eqref{eq:storedEF} is similar to the energy defined by Vandenbosch in~\cite{Vandenbosch2010} for the vacuum case. For simplicity we express the energy using the scalar potential $\phi$ and the vector potential $\Av$ in the Lorentz gauge~\cite{Jackson1999,Landau+Lifshitz1984,vanBladel2007}, so that
$(\nabla^2+k^2)\phi(\rv)=-\rho(\rv)/\epsilon_0$
and $(\nabla^2+k^2)\Av(\rv)=-\mu_0\Jv(\rv)$ and therefore 
\begin{equation}\label{eq:scalarpotential1}	
	\phi(\rv) = \epsilon_0^{-1}(G\ast\rho)(\rv)=\frac{1}{\epsilon_0}\int_V G(\rv-\rv')\rho(\rv')\diffV'
\end{equation}
and
\begin{equation}\label{eq:vectorpotential1}
	\Av(\rv) = \mu_0 (G\ast\Jv)(\rv) =\mu_0\int_V G(\rv-\rv')\Jv(\rv')\diffV',
\end{equation}
where $\ast$ denotes convolution and $G$ is the outgoing Green's function 
\ie $G(\rv)=\eu^{\iu kr}/(4\pi r)$ and $r=|\rv|$. The vector and scalar potentials are related by
$\nabla\cdot\Av=\iu k \phi/\clight$ 
 and the electric and magnetic fields are given by~\cite{Jackson1999}
\begin{equation}
	\Ev = \iu\omega\Av-\nabla\phi
	\qtext{and }
	\Hv =\mu_0^{-1}\nabla\times\Av.
\end{equation}
We also use the corresponding far-field potentials defined by
\begin{equation}\label{eq:scalarpot}
	\phi_{\infty}(\rvh) = \frac{1}{4\pi\epsilon_0}\int_V\rho(\rv')\eu^{-\iu k\rvh\cdot\rv'}\diffV'
\end{equation}
and
\begin{equation}\label{eq:vectorpot}
	\Av_{\infty}(\rvh) = \frac{\mu_0}{4\pi}\int_V\Jv(\rv')\eu^{-\iu k\rvh\cdot\rv'}\diffV'
\end{equation}
giving the electric far-field 
\begin{equation}\label{eq:farfieldinpot}
	\Fv(\rvh) = \iu\omega\Av_{\infty}(\rvh) 
	-\rvh\iu k\phi_{\infty}(\rvh).
\end{equation}
Using that the far-field is orthogonal to $\rvh$, \ie $\rvh\cdot\Fv=0$, the far-field radiation pattern obeys 
\begin{equation}\label{eq:farfieldampinpot}
	|\Fv(\rvh)|^2 = \omega^2|\Av_{\infty}(\rvh)|^2 - k^2|\phi_{\infty}(\rvh)|^2.
\end{equation}
The electric energy density is proportional to
\begin{multline} \label{eq:Eenergydens}
	|\Ev|^2 = \omega^2|\Av|^2-2\Re\{\iu\omega\Av\cdot\nabla\phi^{\ast}\}
	+|\nabla\phi|^2\\
	=\omega^2|\Av|^2
	-2 k^2|\phi|^{2}
	+|\nabla\phi|^2
	-2\Re\{\iu\omega\nabla\cdot(\phi^{\ast}\Av)\},
\end{multline}
where we used $\nabla\cdot(\phi^{\ast}\Av)=\phi^{\ast}\nabla\cdot\Av+\Av\cdot\nabla\phi^{\ast}=\iu k|\phi|^{2}/\clight+\Av\cdot\nabla\phi^{\ast}$. We integrate this result over a large sphere to get the far-field type stored electric energy~\eqref{eq:storedEF} expressed in the potentials
\begin{multline}\label{eq:storedEFinpot}
	\frac{4\WFE}{\epsilon_0} 
	= \int_{\Rr^3}  |\Ev(\rv)|^2-\frac{|\Fv(\rvh)|^2}{r^2} \diffV\\
	=\int_{\Rr^3} |\nabla\phi|^2-k^2|\phi|^{2}
	+\omega^2\left(|\Av|^2-\frac{|\Av_{\infty}|^2}{r^2}\right)
	-k^2\left(|\phi|^2-\frac{|\phi_{\infty}|^2}{r^2}\right)\diffV,
\end{multline}
where we applied the divergence theorem to the integration of the last term in \eqref{eq:Eenergydens}, obtaining via the discussion in \eqref{eq:farfieldinpot} and \eqref{eq:farfieldampinpot} that $\int_{\Omega}\Im\{\phi^{\ast}(r\rvh)A_{\mrm{r}}(r\rvh)\}r^2\diff\Omega\to 0$ as the radius $r\to\infty$ in $\Rr^3$, see~\eqref{eq:farfieldinpot}.

Use the energy identity for the Helmholtz equation, $|\nabla\phi|^2-k^2|\phi|^2 = \epsilon_0^{-1}\Re\{\phi\rho^{\ast}\} + \nabla\cdot(\Re\{\phi^{\ast}\nabla\phi\})$, and that $\phi^*\nabla\phi\rightarrow \iu k \rvh|\phi|^2$ for large enough $r$, to rewrite the first two terms in~\eqref{eq:storedEFinpot} as
\begin{multline}\label{eq:storedEFinpot1}
	\int_{\Rr^3}|\nabla\phi(\rv)|^2-k^2|\phi(\rv)|^{2}\diffV
	=\epsilon_0^{-1}\Re\int_V\phi(\rv)\rho^{\ast}(\rv)\diffV\\
	=\int_V\!\!\int_V \rho(\rv_1)\frac{\cos(k|\rv_1-\rv_2|)}{4\pi\epsilon_0^2 |\rv_1-\rv_2|}\rho^{\ast}(\rv_2)\diffVa\diffVb,
\end{multline}
where we also used that the surface term vanishes.
The Green's function identity, see App.~\ref{sec:GFI}
\begin{equation}\label{eq:GreenId0}
	\int_{\Rr^3} G(\rv-\rv_1)G^{\ast}(\rv-\rv_2)-\frac{\eu^{-\iu k(\rv_1-\rv_2)\cdot\rvh}}{16\pi^2r^2}\diffV
	=-\frac{\sin(kr_{12})}{8\pi k}
+ \iu \frac{r_1^2-r_2^2}{8\pi r_{12}}\jop_1(kr_{12}),
\end{equation}
where $\jop_1(z)=(\sin(z)-z\cos(z))/z^2$ is a spherical Bessel function~\cite{vanBladel2007},
is used to rewrite the two remaining terms in~\eqref{eq:storedEFinpot} as
\begin{multline}\label{eq:storedEFinpot2}
	\int_{\Rr^3}|G\ast\Jv|^2
	-\frac{|\int_V\eu^{-\iu k\rv'\cdot\rvh}\Jv(\rv')\diffV'|^2}{16\pi^2r^2}\diffV
	=-\int_V\!\!\int_V\Jv(\rv_1)\cdot\frac{\sin(k|\rv_1-\rv_2|)}{8\pi k}\Jv^{\ast}(\rv_2)\diffVa\diffVb\\
	+\iu\int_V\!\!\int_V\Jv(\rv_1)\cdot \frac{r_1^2-r_2^2}{8\pi r_{12}}\jop_1(kr_{12})\Jv^{\ast}(\rv_2)\diffVa\diffVb
\end{multline}
and
\begin{multline}\label{eq:storedEFinpot3}
	\int_{\Rr^3}|G\ast\rho|^2
	-\frac{|\int_V\eu^{-\iu k\rv'\cdot\rvh}\rho(\rv')\diffV'|^2}{16\pi^2r^2}\diffV
	=-\int_V\!\!\int_V\rho(\rv_1)\frac{\sin(k|\rv_1-\rv_2|)}{8\pi k}\rho^{\ast}(\rv_2)\diffVa\diffVb\\
	+\iu\int_V\!\!\int_V\rho(\rv_1)\frac{r_1^2-r_2^2}{8\pi r_{12}}\jop_1(kr_{12})\rho^{\ast}(\rv_2)\diffVa\diffVb.
\end{multline}
We note that the first terms in the right-hand side of~\eqref{eq:storedEFinpot2} and~\eqref{eq:storedEFinpot3} only depend on the distance $r_{12}=|\rv_1-\rv_2|$ and are hence coordinate independent, whereas the last terms depend on the coordinate system due to the factor $r_1^2-r_2^2=(\rv_1+\rv_2)\cdot(\rv_1-\rv_2)$. The coordinate dependence originates in the explicit evaluation of the integral in~\eqref{eq:GreenId0} over large spherical volumes $\Rr^3$ that is necessary due to the slow convergence of the integral in~\eqref{eq:GreenId0}, see also App.~\ref{sec:GFI}. 

Collecting the terms in~\eqref{eq:storedEFinpot1}, \eqref{eq:storedEFinpot2}, and~\eqref{eq:storedEFinpot3}, we get a quadratic form in the current density $\Jv$ for the far-field type stored electric energy~\eqref{eq:storedEFinpot} as 
\begin{equation}\label{eq:WeF}
	\WFE = \WE_{\mrm{F_0}} 
	+ W_{\mrm{F_1}} + W_{\mrm{F_2}},
\end{equation}
where $W_{\mrm{F_0}}^{\mrm{(E)}}+ W_{\mrm{F_1}}$ is the coordinate independent part
\begin{multline}\label{eq:WeF01}
	\WE_{\mrm{F_0}}+ W_{\mrm{F_1}}
	=\frac{\mu_0}{4}
	\int_{V}\!\int_{V}\nabla_1\cdot\Jv(\rv_1)\nabla_2\cdot\Jv^{\ast}(\rv_2)
\frac{\cos(kr_{12})}{4\pi k^2 r_{12}} \\	
-\big(k^2\Jv(\rv_1)\cdot\Jv(\rv_2)^{\ast}-\nabla_1\cdot\Jv(\rv_1)\nabla_2\cdot\Jv^{\ast}(\rv_2)\big)
	\frac{\sin(k r_{12})}{8\pi k}	\diffVa\diffVb
\end{multline}
and $W_{\mrm{F_0}}^{\mrm{(E)}}$ and $W_{\mrm{F_1}}$ contains the $\cos$ and $\sin$ parts, respectively.
The coordinate dependent part is
\begin{equation}\label{eq:WemF2}
	W_{\mrm{F_2}}
	=\frac{\mu_0}{4}\int_{V}\!\int_{V}
	\Im\big\{k^2\Jv_1\cdot\Jv_2^{\ast}
	-\nabla_1\cdot\Jv_1\nabla_2\cdot\Jv^{\ast}_2\big\}
	\frac{r_1^2-r_2^2}{8\pi r_{12}}\jop_1(kr_{12})
	\diffVa\diffVb,
\end{equation}
where $\Jv_n=\Jv(\rv_n)$, $n=1,2$.
The coordinate independent part $W_{\mrm{F_0}}^{\mrm{(E)}}+ W_{\mrm{F_1}}$ is identical to the energy by Vandenbosch in~\cite{Vandenbosch2010} for vacuum and hence presents a clear interpretation of the energy~\cite{Vandenbosch2010} in terms of~\eqref{eq:storedEF}. We also see that the definition~\eqref{eq:storedEF} explains the peculiar effects of negative stored energies~\cite{Gustafsson+etal2012a} and suggests a remedy to it in~\eqref{eq:Qbound}. The coordinate dependent part $W_{\mrm{F_2}}$ is more involved. A similar coordinate dependent term is observed in~\cite{Yaghjian+Best2005}.
Obviously the actual stored energy, as any physical quantity, should be independent of the coordinate system. First, we observe that $W_{\mrm{F_2}}=0$ for any current density that has a constant phase. This includes the fields originating from single spherical modes on spherical surfaces and hence most cases in~\cite{Chu1948,Collin+Rothschild1964,Thal2006,Hansen+Collin2009}. It also includes currents in the form of single characteristic modes~\cite{Capek+etal2012}. We also get the coordinate independent part by taking the average of the stored energy from $\Jv$ and $\Jv^{\ast}$. The term $W_{\mrm{F_2}}$ is further analyzed in Secs~\ref{S:coordinatedependent} and~\ref{S:small}.

For the stored magnetic energy we can use $|\Bv|^2=|\nabla\times\Av|^2$ or simpler the energy identity~\eqref{eq:energyid_Im}, to directly get the difference
\begin{equation}
	\int_{\Rr^3}\mu_0|\Hv(\rv)|^2-\epsilon_0|\Ev(\rv)|^2\diffV
	=\Re\int_V\Av(\rv)\cdot\Jv^{\ast}(\rv)-\phi(\rv)\rho^{\ast}(\rv)\diffV,
\end{equation}
where we used
\begin{equation}
	\Ev\cdot\Jv^{\ast}
	=\iu\omega\Av\cdot\Jv^{\ast}-\nabla\cdot(\phi\Jv^{\ast})-\iu\omega\phi\rho^{\ast}.
\end{equation}

This gives the far-field type stored magnetic energy $W^{\mrm{(M)}}_{\mrm{F}}=W^{\mrm{(M)}}_{\mrm{F_0}}+W_{\mrm{F_1}}+W_{\mrm{F_2}}$, where the coordinate independent part
\begin{multline}\label{eq:WmF01}
	\WM_{\mrm{F_0}}+W_{\mrm{F_1}}=
	\frac{\mu_0}{4}\int_{V}\!\int_{V}\Jv(\rv_1)\cdot\Jv^{\ast}(\rv_2)\frac{\cos(kr_{12})}{4\pi r_{12}} \\
	-\big(k^2\Jv(\rv_1)\cdot\Jv(\rv_2)^{\ast} 
	-\nabla_1\cdot\Jv(\rv_1)\nabla_2\cdot\Jv^{\ast}(\rv_2)\big)
	\frac{\sin(kr_{12})}{8\pi k}\diffVa\diffVb
\end{multline}
is expressed as a quadratic form in $\Jv$, see also~\cite{Vandenbosch2010}. We also have the radiated power
\begin{equation}
	P_{\mrm{r}}=\frac{\eta_0}{2k}\int_{V}\!\int_{V}
	\big(k^2\Jv(\rv_1)\cdot\Jv^{\ast}(\rv_2) 
	-\nabla_1\cdot\Jv(\rv_1)\nabla_2\cdot\Jv^{\ast}(\rv_2)\big)
	\frac{\sin(kr_{12})}{4\pi r_{12}}
	\diffVa\diffVb.		
\end{equation}

It is illustrative to rewrite the coordinate independent far-field stored energy in the potentials:
\begin{equation}
	\WE_{\mrm{F_0}}
	=\frac{1}{4}\Re\int_V\rho^{\ast}\phi\diffV,\quad
	\WM_{\mrm{F_0}}
	= \frac{1}{4}\Re\int_V \Jv^{\ast}\cdot\Av\diffV.
\end{equation}
We note that the sum of the first terms, $\WE_{\mrm{F_0}}+\WM_{\mrm{F_0}}$, corresponds to a frequency-domain version of the energy expression by Carpenter~\cite{Carpenter1989}, see also~\cite{Endean+Carpenter1992,Uehara+etal1992}. Moreover, they reduce to well-known electrostatic and magnetostatic expressions in the low-frequency limit~\cite{Jackson1999}. 

We follow standard notation in the method of moments (MoM) and introduce the operators $\Lop_\mrm{e}$ and $\Lop_\mrm{m}$ such that $\Lop=\Lop_\mrm{e}-\Lop_\mrm{m}$ is the integral operator associated with the electric field integral equation (EFIE)~\cite{Jin2010}. Here, the operators are generalized to volumes and defined from
\begin{equation}\label{eq:ZeJ}
	\langle \Jv,\Lop_\mrm{e}\Jv\rangle
 	=\frac{-1}{\iu k} \int_{V}\!\int_{V}\nabla_1\cdot\Jv(\rv_1)\nabla_2\cdot\Jv^{\ast}(\rv_2) G(\rv_1-\rv_2)
	\diffVa\diffVb,
\end{equation}
\begin{equation}\label{eq:ZmJ}
	\langle \Jv,\Lop_\mrm{m}\Jv\rangle
	=\iu k\int_{V}\!\int_{V}\Jv(\rv_1)\cdot\Jv^{\ast}(\rv_2)G(\rv_1-\rv_2)\diffVa\diffVb,
\end{equation}
and
\begin{multline}\label{eq:ZemJ}
	\langle \Jv,\Lop_\mrm{em}\Jv\rangle
	=\frac{\iu k}{2}\int_{V}\!\int_{V}
	\Big(\frac{1}{k}\nabla_1\cdot\Jv(\rv_1)\nabla_2\cdot\Jv^{\ast}(\rv_2) \\
	-k\Jv(\rv_1)\cdot\Jv^{\ast}(\rv_2)\Big)
	\partder{G(\rv_1-\rv_2)}{k}\diffVa\diffVb.
\end{multline}
They are defined such that the stored electric and magnetic energies and radiated power are
	\begin{align}
		\WE_{\mrm{F_0}}&=\frac{\eta_0}{4\omega}\Im\langle \Jv,\Lop_\mrm{e}\Jv\rangle \label{eq:ZtoWE}\\
		\WM_{\mrm{F_0}}&=\frac{\eta_0}{4\omega}\Im\langle \Jv,\Lop_\mrm{m}\Jv\rangle \label{eq:ZtoWM}\\
		W_{\mrm{F_1}}&=\frac{\eta_0}{4\omega}\Im\langle \Jv,\Lop_\mrm{em}\Jv\rangle \label{eq:ZtoWEM}\\
		P_{\mrm{r}}&=\frac{\eta_0}{2}\Re\langle \Jv,(\Lop_\mrm{e}-\Lop_\mrm{m})\Jv\rangle. \label{eq:ZtoP}
	\end{align} 
Efficient evaluation of the $\Lop$ operator is instrumental in MoM implementations where the discretized versions are often referred to as impedance matrices. The relations above show that the corresponding matrices for the coordinate independent stored and radiated energies are available by evaluating the real and imaginary parts of the MoM impedance matrices with the addition of the mixed part~\eqref{eq:ZemJ}.
The stored energy is hence computed with negligible additional computational cost in MoM implementations. Moreover, \eqref{eq:ZtoP} shows that $\Re\Lop$ is positive semidefinite.

\section{Coordinate dependent term}\label{S:coordinatedependent} 
The stored electric~\eqref{eq:WeF} and magnetic energies contain the potentially coordinate dependent part $W_{\mrm{F_2}}$ defined in~\eqref{eq:WemF2}.
Assume that $W_{\mrm{F_2}}=W_{\mrm{F_2},\myvec{0}}$ for one coordinate system. Consider a shift of the coordinate system $\rv\to\dv+\rv$ and use that $r_1^2-r_2^2\to r_1^2-r_2^2 + 2\dv\cdot(\rv_1-\rv_2)$. This gives the coordinate dependent term
\begin{equation}
	W_{\mrm{F_2},\dv}
	=W_{\mrm{F_2},\myvec{0}}
	+k\dv\cdot \Wv,
\end{equation}
where $\Wv = \Wv_{\!\rho}+\Wv_{\!\mrm{J}}$ and
\begin{multline}
	\Wv_{\!\rho} = \frac{\iu}{2\epsilon_0}\int_V\int_V\rho(\rv_1)\nabla_1\frac{\sin(kr_{12})}{8\pi kr_{12}}\rho^{\ast}(\rv_2)\diffVa\diffVb\\
	=\frac{k\epsilon_0}{4} \int_{\Omega}\rvh\Big|\int_V \frac{\rho(\rv)\eu^{-\iu k\rvh\cdot\rv}}{4\pi\epsilon_0}\diffV\Big|^2\diff\Omega
	=\frac{k\epsilon_0}{4}\int_{\Omega}|\phi_{\infty}(\rvh)|^2\rvh\diff\Omega
\end{multline}
and we used~\eqref{eq:GreenFId1}, the identity
\begin{equation}
	\nabla_1\frac{\sin(kr_{12})}{4\pi kr_{12}}
	=-\iu k\lim_{r\to\infty}\int_{|\rv|=r}\rvh G(\rv-\rv_1)G^{\ast}(\rv-\rv_2)\diffS
	=\frac{-\iu k}{16\pi^2}\int_{\Omega}\rvh \eu^{-\iu k\rvh\cdot(\rv_1-\rv_2)}\diff\Omega,
\end{equation}
and the far-field potential~\eqref{eq:scalarpot}.
Similarly, the current part is
\begin{equation}
	\Wv_{\!\mrm{J}} = -\frac{\iu\mu_0}{2}\int_V\!\int_V\Jv(\rv_1)\cdot\Jv^{\ast}(\rv_2)\nabla_1\frac{\sin(kr_{12})}{8\pi kr_{12}}\diffV_{\!1}\diffV_{\!2}
	= -\frac{k}{4\mu_0} \int_{\Omega}|\Av_{\infty}(\rvh)|^2\rvh\diff\Omega.
\end{equation}
Using the far-field identity~\eqref{eq:farfieldampinpot} gives $\Wv$ as
\begin{equation}
	\Wv = -\frac{\epsilon_0}{4k}\int_{\Omega}|\Fv(\rvh)|^2\rvh\diff\Omega.
\end{equation}
The corresponding Q factor is hence shifted as
\begin{equation}
	\varDelta Q_{\mrm{F_2}}=\frac{-k\dv\cdot\int_{\Omega}\rvh|\Fv(\rvh)|^2\diff\Omega}{2 \int_{\Omega}|\Fv(\rvh)|^2\diff\Omega},	
\end{equation}
where we see that $|\varDelta Q_{\mrm{F_2}}|\leq ka$ for all coordinate shifts within the smallest circumscribing sphere, see Fig.~\ref{fig:objectgeo}. We note that this term is similar to the coordinate dependence observed in~\cite{Yaghjian+Best2005}.

Consider a spherical current sheet to illustrate the coordinate dependence. Let the far field be $\Fv\sim \alpha_1\Yvop_{1\mrm{e}01}+\alpha_2\Yvop_{2\mrm{o}11}$, \ie a combination of a $\zvh$ directed magnetic dipole and a $\yvh$ directed electric dipole, see App.~\ref{S:sphwaves}. This gives the shift $\varDelta Q_{\mrm{F_2}}=-kx/4$. We also have $Q_{\mrm{F_2},\myvec{0}}=0$ for the case of a coordinate system centered in the sphere as $r_1=r_2$ gives $Q_{\mrm{F_2},\dv}=-kx/4$, where $x=\dv\cdot\xvh$ and $\dv$ is the vector to the center of the sphere.

\section{Small structures}\label{S:small} 
Evaluation of the stored energy for antenna Q is most interesting for small structures, where $Q$ is large, \eg $Q\geq 10$, and can be used to quantify the bandwidth of antennas~\cite{Chu1948,Yaghjian+Best2005,Gustafsson+Nordebo2006b,Vandenbosch2011,Gustafsson+etal2012a}. The low-frequency expansion of the stored energy are presented in~\cite{Geyi2003,Vandenbosch2010,Vandenbosch2011,Gustafsson+etal2012a}. Here, we base it on the low-frequency expansion $\Jv=\Jv^{\mrm{(0)}}+k\Jv^{\mrm{(1)}}+\Ordo(k^2)$ as $k\to 0$, where $\nabla\cdot\Jv^{\mrm{(0)}}=0$ and the static terms $\Jv^{\mrm{(0)}}$ and $\rho_0=-\iu\nabla\cdot\Jv^{\mrm{(1)}}/\clight$ have a constant phase. 
For the corresponding asymptotic expansions of the Q-factor components in~\eqref{eq:WeF}, we note that the coordinate dependent part vanishes if $\Jv$ and $\rho(\rv)$ have constant phase. This gives 
\begin{multline}
	\Im\{\rho(\rv_1)\rho^{\ast}(\rv_2)\}
	=\Im\{(\rho_0(\rv_1)+k\rho_1(\rv_1))(\rho_0^{\ast}(\rv_2)+k\rho_1^{\ast}(\rv_2))\}+\Ordo(k^2)\\
	=k\Im\{\rho_0(\rv_1)\rho_1^{\ast}(\rv_2)+\rho_1(\rv_1)\rho_0^{\ast}(\rv_2)\}+\Ordo(k^2)
\end{multline}
as $k\to 0$ and similarly for $\Jv$. 
The different parts of the stored energy~\eqref{eq:WeF} contribute to the Q-factor asymptotically
\begin{equation}\label{eq:QEMexp}
	Q^{\mrm{(E,M)}}_{\mrm{F_0}} \sim \frac{1}{(ka)^3},\ 
	Q_{\mrm{F_1}} \sim \frac{1}{ka},\ 
	Q_{\mrm{F_2}} \sim ka
\end{equation}
as $ka\to 0$, where $a$ is the radius of smallest circumscribing sphere and the coordinate system is centered inside the sphere.

We can compare the expansion~\eqref{eq:QEMexp} with the Chu bound~\cite{Chu1948}
\begin{equation}
	Q_{\mrm{Chu}} = \frac{1}{(ka)^3} + \frac{1}{ka},
\end{equation}
where it is seen that $Q_{\mrm{Chu}}$ has components that are of the same order as $Q^{\mrm{(E,M)}}_{\mrm{F_0}}$ and $Q_{\mrm{F_1}}$ and hence that these terms are essential to produce reliable results. This is also the conclusion from Sec.~\ref{S:Sphere} in~\eqref{eq:QpE0}, where it is shown that the Q-factors differ by $ka$.

The coordinate dependent part $Q_{\mrm{F_2}}$ is negligible for small structures and is of the same order as the difference between the far-field~\eqref{eq:storedEF} and power~\eqref{eq:storedEP} type as seen by the bound~\eqref{eq:Qbound}. We also note that the importance of $Q$ diminishes as $Q$ approaches unity. This also restricts the interest of the results to small antennas. 

\section{Stored energy from the input impedance}\label{S:Zenergy}
The bandwidth of an antenna is often determined from the antenna input impedance. The fractional bandwidth is related to the Q-value for simple lumped resonance circuits~\cite{Yaghjian+Best2005}
\begin{equation}\label{eq:Q2Bandwidth}
	B \approx \frac{2\refl_0}{Q\sqrt{1-\refl_0^2}},
\end{equation}
where $\refl_0$ is the threshold of the reflection coefficient. The Fano limit~\cite{Fano1950,Gustafsson+Nordebo2006b} for a resonant circuit, $B\leq 27.29/(Q|\refl_{0,\dB}|)$, can be used for the bandwidth after matching, where $\refl_{0,\dB}=20\log_{10}\refl_0$.
For more general circuits we consider the Q values determined from the differentiated input impedance and from the stored energy in equivalent circuit models. 

The Q factors from the \textit{differentiated impedance} at the resonance angular frequency $\omega_0$ is~\cite{Yaghjian+Best2005,Gustafsson+Nordebo2006b} 
\begin{equation}\label{eq:QZp}
	\QZp(\omega_0)=\frac{\omega_0 |Z_{\mrm{m}}'|_{\omega=\omega_0}}{2R(\omega_0)}
	=\omega_0|\refl'|_{\omega=\omega_0},
\end{equation}
where ${}'$ denotes differentiation with respect to $\omega$ and $Z_{\mrm{m}}$ is the input impedance $Z=R+\ju X$, with $\ju=-\iu$, tuned to resonance with a lumped series (or analogous for lumped elements in parallel) inductor or capacitor
\begin{equation}
	Z_{\mrm{m}}(\omega) = Z(\omega) -
	\begin{cases}
		\ju X(\omega_0)\omega/\omega_0&\text{if } X(\omega_0)<0\\
		\ju X(\omega_0)\omega_0/\omega&\text{if } X(\omega_0)>0.
	\end{cases}
\end{equation}
In addition to the Q factor in~\eqref{eq:QZp}, we determine the stored energy in the lumped element normalized with the radiated power as
\begin{equation}
	\varDelta \QZp(\omega_0) = \frac{|X(\omega_0)|}{R(\omega_0)}
\end{equation}
giving the electric and magnetic Q factors
\begin{equation}\label{eq:QZE}
	 \QZpE = 
	\begin{cases}
		\QZp &\text{if } X(\omega_0)<0\\
		\QZp-\varDelta \QZp &\text{if } X(\omega_0)>0
	\end{cases}	
\end{equation}
and
\begin{equation}\label{eq:QZM}
	\QZpM = 
	\begin{cases}
		\QZp &\text{if } X(\omega_0)>0\\
		\QZp-\varDelta \QZp &\text{if } X(\omega_0)<0,
	\end{cases}	
\end{equation}
respectively. 

The Q factor can alternatively be determined from the stored and dissipated energy in an equivalent circuit model for the input impedance. The input impedance of small antennas can often be approximated with simple resonance circuits.
More accurate circuit models can be synthesized using \eg Brune, Bott and Duffin, Miyata, or Darlington synthesis~\cite{Brune1931,Wing2008}. The synthesis methods can produce different circuit topologies so the circuits are not unique. It is also possible to synthesize lumped circuits with an internal stored energy that is non-observable from the input impedance. 

Here, we consider a rational approximation of the input impedance for the antenna. In the range $\omega_1\leq\omega\leq\omega_2$ we use a rational function of order $(m_1,m_2)$, with $|m_1-m_2|\leq 1$, that is fitted to the input impedance using the MATLAB function \verb+invfreqs+. The order is chosen as low as possible such that the relative error is below some threshold level, here $10^{-3}$, and that the rational function is a positive real (PR) function~\cite{Wing2008}. We use Brune synthesis~\cite{Brune1931,Wing2008} to construct an equivalent circuit model and determine the stored and dissipated energy, \ie
\begin{equation}\label{eq:circuitQ}
	\QBE = \frac{\sum_n |I_n|^2/C_n}{\omega\sum_n R_n|I_n|^2}
	\qtext{and }
	\QBM = \frac{\omega\sum_n L_n|I_n|^2}{\sum_n R_n|I_n|^2},
\end{equation}
where $C_n$, $L_n$, $R_n$, and $I_n$ are the capacitance, inductance, resistance and current in branch $n$, see also App.~\ref{S:Brune}.

\section{Examples}\label{S:antennaex}
To interpret the different proposals for stored electromagnetic energy, we consider analytic and numerical examples. The first analytic example illustrates the relation between the stored energy defined by subtraction of the power flow~\eqref{eq:storedEP}, used in~\cite{Chu1948,Collin+Rothschild1964,McLean1996}, and the far-field power~\eqref{eq:storedEF} similar to~\cite{Vandenbosch2010} for spherical modes and shows that their Q factors differ by $ka$. The second example compares the Q factors and the associated bandwidths defined from the input impedance using Brune synthesis~\cite{Brune1931} and differentiation for a lumped circuit network. Finally, we consider dipole, loop, and inverted-L antennas to compare the Q factors from the input impedance using Brune synthesis~\cite{Brune1931} and differentiation~\cite{Yaghjian+Best2005} with the stored energy determined from the current density~\eqref{eq:WeF01}~\cite{Vandenbosch2010}. We illustrate both the Q factors and the resulting bandwidth after matching.   

\subsection{Numerical example for spherical shells}
\begin{figure}[t]
\begin{center}
\noindent
  \includegraphics[width=0.98\textwidth]{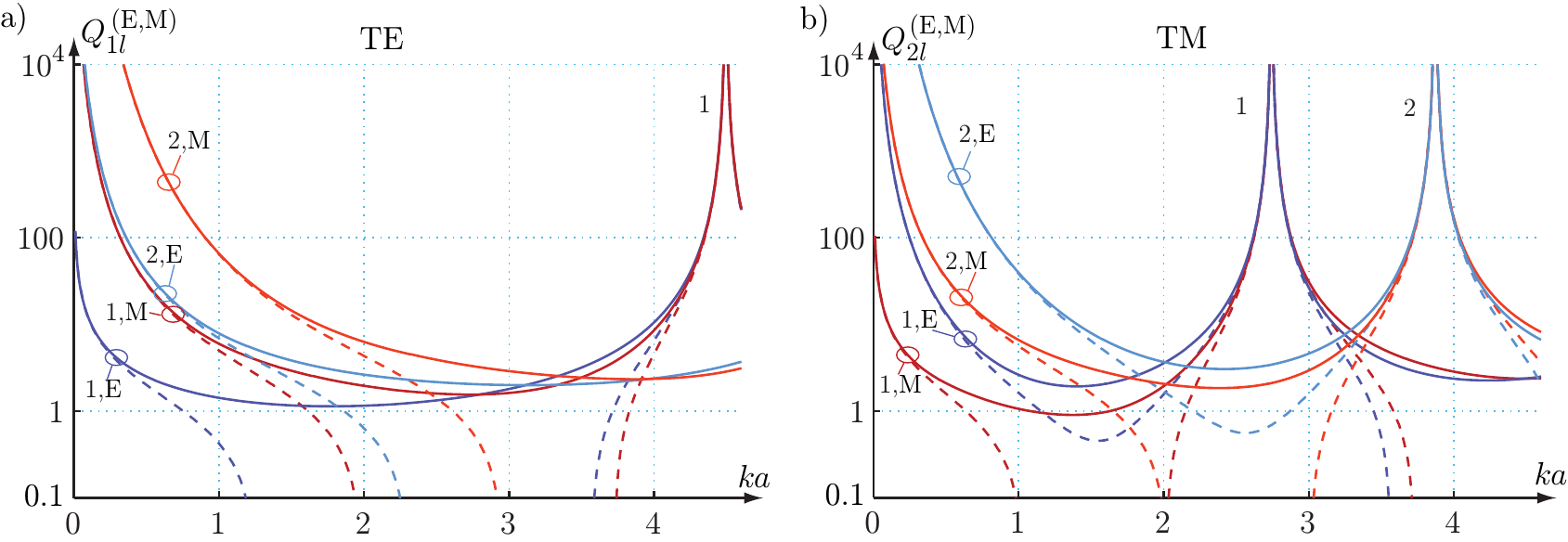}
  \caption{Electric and magnetic Q factors for electrical surface currents $\Jv(\rv)=J_0\Yvop_{\tau\sigma ml}(\rvh)\delta(r-a)$ for  $l=1,2$. Power (solid curves) and far-field (dashed curves) stored energies. They differ by $ka$~\eqref{eq:QpE0}. a) TE ($\tau=1$) modes. b) TM ($\tau=2$) modes.}
  \label{fig:Qsph}
\end{center}
\end{figure}

The two formulations~\eqref{eq:storedEP} and~\eqref{eq:storedEF} for the stored energy can be compared for electric surface currents on spherical shells. This is the case analyzed by Thal~\cite{Thal2006} and Hansen \& Collin~\cite{Hansen+Collin2009}, see also~\cite{Hansen+etal2012} for the case with electric and magnetic surface currents.
We expand the surface current on a sphere with radius $a$ in vector spherical harmonics $\Yvop$, see App.~\ref{S:sphwaves}.
The electric and magnetic Q factors are 
\begin{equation}\label{eq:QiE0}
	Q^{\mrm{(E)}}_{\tau l,\mrm{F}}(\kappa)  
	=-\frac{\big(\kappa\Rop_{\tau l}^{(1)}(\kappa)\Rop_{\tau l}^{(2)}(\kappa)\big)'}{2 (\Rop_{\tau l}^{(1)}(\kappa))^2}
\end{equation}
and
\begin{equation}\label{eq:QiM0}
	Q^{\mrm{(M)}}_{\tau l,\mrm{F}} 
	=Q^{\mrm{(E)}}_{\tau l,\mrm{F}}(\kappa) 
	-\frac{\Rop_{\tau l}^{(2)}(\kappa)}{\Rop_{\tau l}^{(1)}(\kappa)},
\end{equation}
respectively. We note that the expressions for the TE and TM are written in identical forms by using the radial functions~\cite{Hansen1988}, see also~\eqref{eq:radialfunctions}.

The corresponding power flow stored energy is
\begin{equation}\label{eq:QpE0}
	Q^{\mrm{(E,M)}}_{\mrm{\tau l,P}}(\kappa)
	=\frac{2\omega W^{\mrm{(E,M)}}_{\mrm{P}}(\kappa)}{P_{\mrm{r}}(\kappa)}
	=\kappa+Q^{\mrm{(E,M)}}_{\mrm{\tau l,F}}(\kappa),
\end{equation}
where $Q^{\mrm{(E,M)}}_{\mrm{\tau l,F}}$ denotes the electric and magnetic far-field type Q factors in~\eqref{eq:QiE0} and~\eqref{eq:QiM0}.
The difference $\kappa=ka$ is consistent with the interpretation of a standing wave in the interior of the sphere, \cf~\eqref{eq:Qbound}. Moreover, the expressions~\eqref{eq:QiE0} and~\eqref{eq:QiM0} unifies the TE and TM cases and offer an alternative to the expressions in~\cite{Hansen+Collin2009}, here we also note a misprint in (6) in~\cite{Hansen+Collin2009}.

The electric and magnetic Q-factors are depicted in Fig.~\ref{fig:Qsph} for $l=1,2$. The relative differences are negligible for small $ka$ where $Q$ is large. For larger $ka$, where $Q$ can be small, the relative difference is significant although the absolute difference is exactly $ka$. We also note that the $Q$ factors oscillate and can be significant even for large $ka$. This is mainly due to small values of $\Rop^{(1)}_{\tau l}(ka)$ that can be interpreted as a negligible radiated power. Moreover, the Q-factors related to the far-field type stored energy~\eqref{eq:storedEF} is negative in some frequency bands. The corresponding Q-factors related to~\eqref{eq:storedEP} are always non-negative. Moreover, it is observed that $Q^{\mrm{(M)}}_{\mrm{1l,P}}\geq Q^{\mrm{(E)}}_{\mrm{1l,P}}$ for low $ka$ but has regions with $Q^{\mrm{(M)}}_{\mrm{1l,P}}<Q^{\mrm{(E)}}_{\mrm{1l,P}}$ for larger $ka$.

\begin{figure}[t]
\begin{center}
\noindent
  \includegraphics[width=0.45\textwidth]{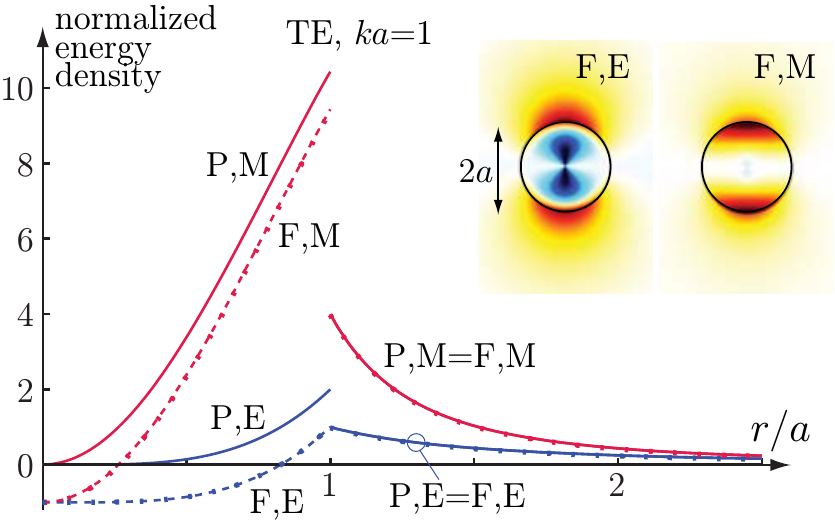}
  \caption{Illustration of the stored electric (E) and magnetic (M) energy densities for the TE ($\tau=1$) mode generated by currents on a spherical shell with radius $ka=1$. Power (solid curves-P) and far-field (dashed curves-F) stored energies. The energy densities are normalized with the radiated power and integrated over spherical shells to emphasize the radial dependence. The angular distribution is also depicted.}
  \label{fig:WemSphere}
\end{center}
\end{figure}

To further analyze the negative values of~\eqref{eq:storedEF}, we depict the stored electric and magnetic energy density over spherical shells related to~\eqref{eq:storedEF} in Fig.~\ref{fig:WemSphere} for a TE spherical current sheet with radius $ka=1$ and a coordinate system with origin at the center of the sphere. The results confirm that the far-field~\eqref{eq:storedEF} and power flow~\eqref{eq:storedEP} stored energy densities are identical outside the sphere. We also see that the far-field stored energy density is negative in parts of the interior region of the sphere, $r<a$, whereas the power flow stored energy density is non-negative.
Moreover, the stored energy density is discontinues at $r=a$ except for the far-field type stored electric energy. The continuity of the far-field type stored electric energy is consistent with the boundary condition that states that tangential components of the electric field are continuous.

\subsection{Resonance circuit}\label{S:ex_resonancecircuit}
\begin{figure}[t]
\begin{center}
  \includegraphics[width=0.45\textwidth]{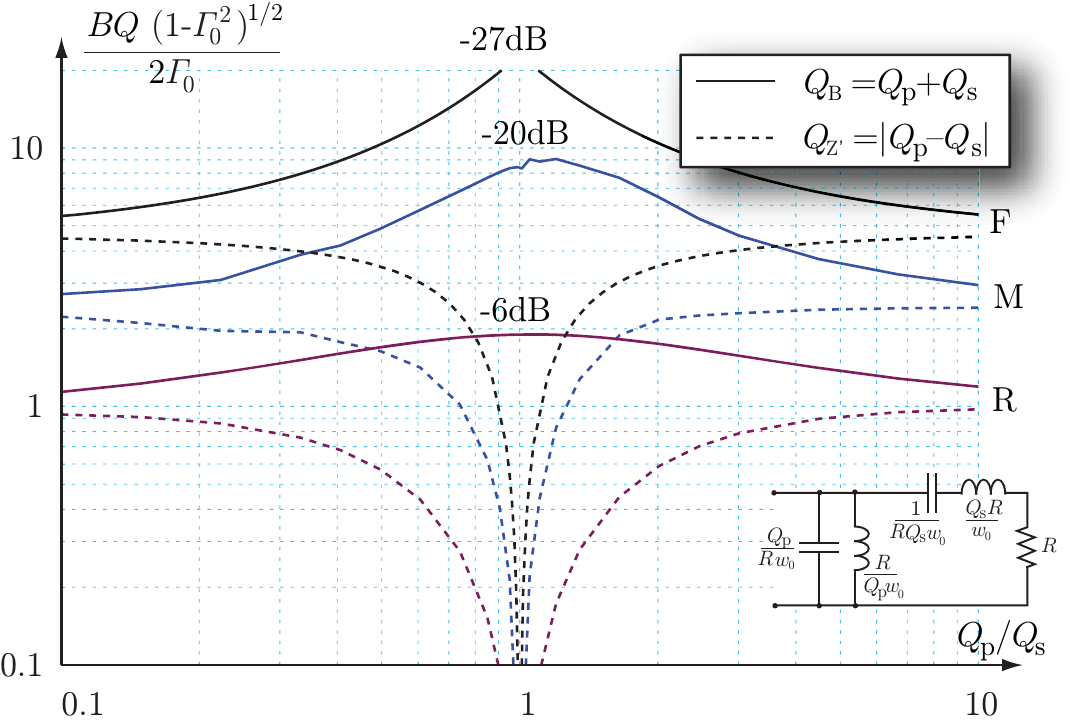}	
	\hspace{5mm}
	\includegraphics[width=0.45\textwidth]{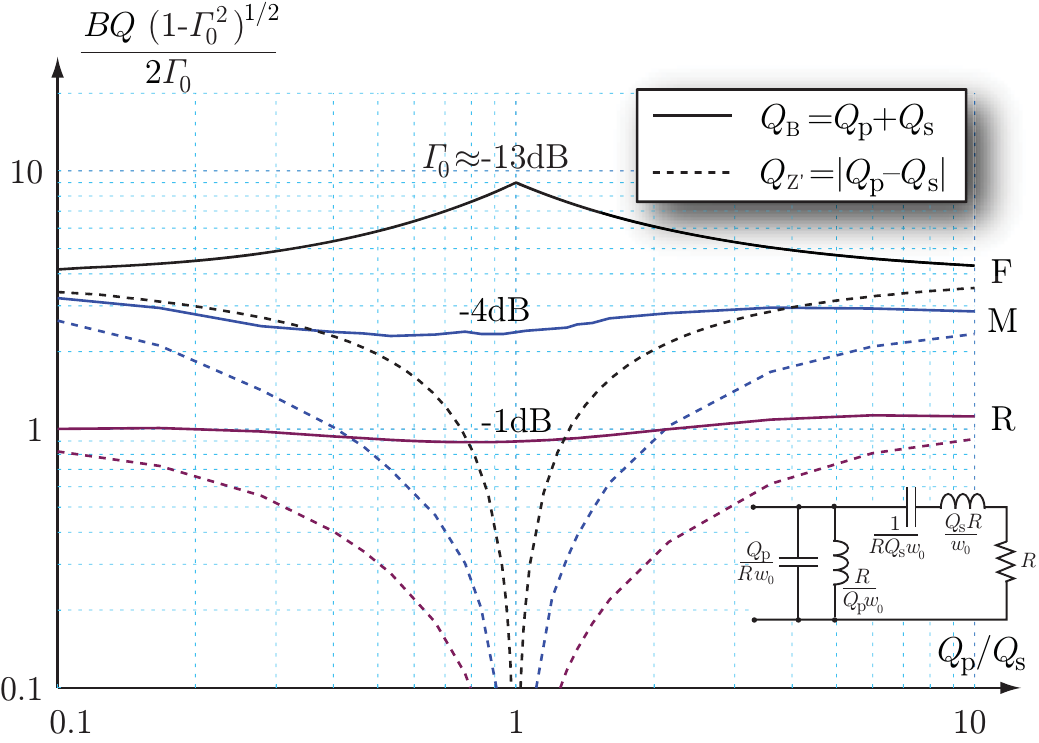}
\end{center}	
  \caption{Illustration of the Q factor fractional bandwidth product for a cascaded shunt LC and series LC network. The Q factors are determined from the stored energies in the circuit elements and from differentiation of the impedance~\eqref{eq:QZp}. The maximal reflection coefficient is determined over the fractional bandwidths $B=\{2,4\}/\QB$ with and without matching networks for the case $Q_{\mrm{s}}=10$ and $1\leq Q_{\mrm{p}}\leq 100$. The matching network is determined using genetic algorithms with four ideal lumped elements. a) $B=2/\QB$ and b) $B=4/\QB$.}
  \label{fig:BQ4pLCsLCR}
\end{figure}

We consider first a simple resonance circuit composed of cascaded shunt LC and series LC networks, see Fig.~\ref{fig:BQ4pLCsLCR}. The elements are chosen to have the same resonance frequency, $\omega_0$, and the element values are expressed in the series and parallel Q factors, $Q_{\mrm{s}}$ and $Q_{\mrm{p}}$, respectively.
The Q from the stored energy, $\QB$, in the circuit elements and differentiation of the input impedance, $\QZp$, are~\cite{Gustafsson+Nordebo2006b}.
\begin{equation}\label{eq:RCLflatmatch}
	\QB = Q_{\mrm{s}} + Q_{\mrm{p}}
	\qtext{and }
	\QZp= |Q_{\mrm{s}} - Q_{\mrm{p}}|,
\end{equation}
respectively. Here, we note that $\QB\geq \QZp$ and that $\QZp=0$ for the case of a flat match $Q_{\mrm{s}} = Q_{\mrm{p}}$. 

The bandwidth and the threshold of the reflection coefficient are related as seen for the simple RCL resonance circuit in~\eqref{eq:Q2Bandwidth}.  
We illustrate the relation between Q and fractional bandwidth $B$, by plotting
	$BQ\sqrt{1-\refl_0^2}/(2\refl_0)$
for $Q=\{\QB,\QZp\}$, \ie $Q$ given by the stored energy in the circuit elements and by differentiation of the impedance in~\eqref{eq:RCLflatmatch}, see Fig.~\ref{fig:BQ4pLCsLCR}. The used scaling removes the ambiguity between $B$ and $\refl$ for a resonance circuit. Moreover, results close to unity means that the relation~\eqref{eq:Q2Bandwidth} holds approximately for the used $Q$. 

We can determine that bandwidth for a given threshold, $\refl_0$ or vice versa. Here, we consider the resulting threshold for the fractional bandwidths $B=2/\QB$ and $B=4/\QB$ to illustrate the dependence on $\QB$ and $\QZp$. This corresponds to well matched but narrow bandwidth and less well matched and wider bandwidth cases. The series Q value is fixed $Q_{\mrm{s}}=10$ whereas the parallel Q is $1\leq Q_{\mrm{p}}\leq 100$. The curves labeled (R) show the bandwidth Q factor product~\eqref{eq:Q2Bandwidth} for the fractional bandwidths $B=2/\QB$ and $B=4/\QB$ using the characteristic impedance $R$ without additional matching networks. 

The product is close to unity at the end points $Q_{\mrm{p}}=\{1,100\}$, where the input impedance resembles a series and parallel RCL circuit. In the region $Q_{\mrm{p}}\approx Q_{\mrm{s}}$ the curves deviates from unity as the input impedance do not resemble an RCL resonance circuit. We also note that the approximation with $\QZp=|Q_{\mrm{s}} - Q_{\mrm{p}}|$ gives vanishing small values showing that the $\QZp$ approximation fails for this case~\cite{Gustafsson+Nordebo2006b}. The use of the Q from the stored energy gives better results. In particular for the wider bandwidth case $B=4/\QB$.

We also consider the case with matching circuits. The Bode-Fano matching limits~\cite{Fano1950,Gustafsson+Nordebo2006b} are depicted by the curves labeled (F) for the cases for $B=2/\QB$ and $B=4/\QB$, see App.~\ref{S:BodeFano}. We note that the Bode-Fano limit mainly depends on the maximal Q value that also can be interpreted as the mean $\Qa=(\QB+\QZp)/2=\max\{Q_{\mrm{p}},Q_{\mrm{s}}\}$.
We use optimization to synthesize lossless matching networks. The curves labeled (M) in Fig.~\ref{fig:BQ4pLCsLCR} show the resulting bandwidth Q factor product~\eqref{eq:Q2Bandwidth} after matching. The first case, $B=2/\QB$, gives a matching threshold $\refl_0$ in the range $-15\dB$ to $-20\dB$ whereas the second case, $B=4/\QB$. gives $\refl_0$ in the range $-3\dB$ to $-6\dB$. 
We consider up to two capacitors and two inductors in the matching network and use a genetic algorithm~\cite{Rahmat-Samii+Michielssen1999} to determine the parameter values.

The results show that the inverse proportionality between $B$ and $Q$ in~\eqref{eq:Q2Bandwidth} is valid for the resonance circuit case far away from $Q_{\mrm{s}} = Q_{\mrm{p}}$. Closer to $Q_{\mrm{s}} = Q_{\mrm{p}}$, the results are better for the stored energy, $\QB=Q_{\mrm{s}}+Q_{\mrm{p}}$, than for the differentiated impedance $\QZp=|Q_{\mrm{s}} - Q_{\mrm{p}}|$. The addition of a matching network increases the bandwidth. Also for this case, the stored energy results are better although they underestimate the bandwidth with up to approximately a factor of two. It should also be noted that the addition of the matching network increases the stored energy and hence the $\QB$, so the $\QB$ after matching can underestimate the bandwidth even more. 

\subsection{Strip dipole}
\begin{figure}[t]
\begin{center}
	\includegraphics[width=0.9\linewidth]{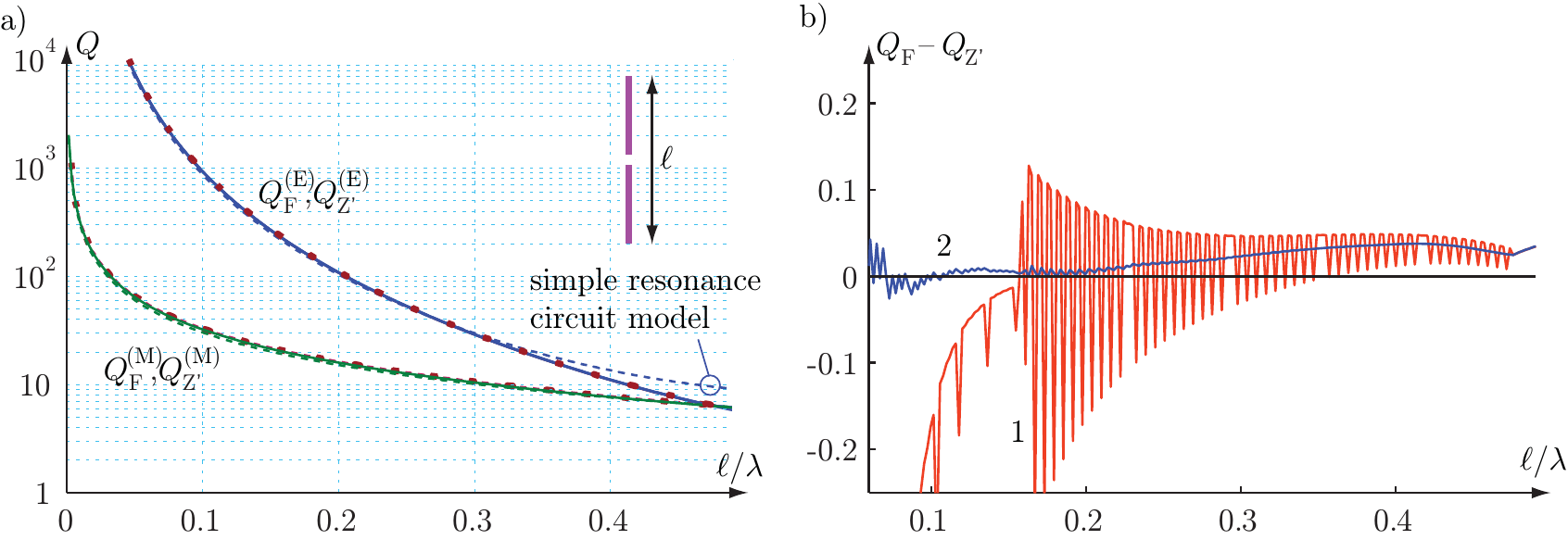}
\end{center}	
  \caption{Illustration of the Q factor for a center feed strip dipole with length $\ell$ and width $\ell/100$. The Q factors are determined from the stored energies~\eqref{eq:WeF01} and~\eqref{eq:WmF01} and from differentiation of the impedance~\eqref{eq:QZE} and~\eqref{eq:QZM}. a) electric and magnetic Q-factors from~\eqref{eq:WeF01},~\eqref{eq:WmF01}, the circuit model (dashed curves), and differentiation of the impedance $\QZp$. b) difference between the computed Q-factors $\QF-\QZp$, where $\QZp$ is computed from a difference scheme and analytic differentiation of a high order rational approximation in 1 and 2, respectively.}
  \label{fig:DipoleQcomp}
\end{figure} 
Consider a center fed strip dipole with length $\ell$ and width $\ell/100$ modeled as perfectly electric conducting (PEC). The Q-factors~\eqref{eq:QEQM} determined from the integral expressions $\QFE$ in~\eqref{eq:WeF01} and $\QFM$ in~\eqref{eq:WmF01}, the simple resonance circuit model~\cite{Gustafsson2010b}, and differentiation of the impedance~\cite{Yaghjian+Best2005,Gustafsson+Nordebo2006b} are compared in Fig.~\ref{fig:DipoleQcomp}a. The circuit model is based on the circuit representations of the lowest order spherical modes~\cite{Thal1978} with the lumped elements determined with the approach in~\cite{Gustafsson2010b}. The Q-factors from the simple resonance circuit model approximates the integral expression very well for $\ell<0.3\lambda$ but starts to differ for shorter wavelengths where the circuit model is less accurate, see Fig.~\ref{fig:DipoleQcomp}a. The difference $\QF-\QZp$ is also depicted in Fig.~\ref{fig:DipoleQcomp}b. We see that the difference is negligible for the considered wavelengths. Curve (1) shows $\QZp$ computed with a finite difference scheme. The curve is sensitive to noise and the used discretization. The noise is suppressed by approximating the impedance with a high order polynomial and performing analytic differentiation as seen by curve (2). 

\begin{figure}[t]
\begin{center}
  \includegraphics[width=0.98\linewidth]{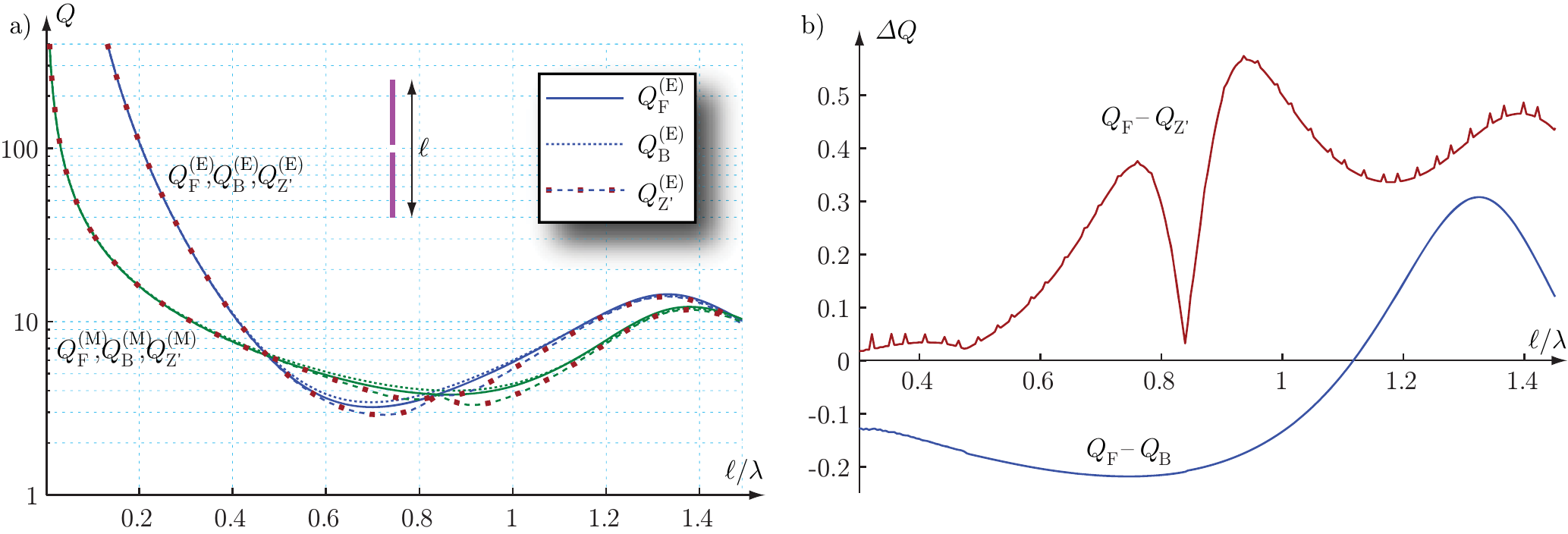}
\end{center}	
  \caption{Illustration of the Q factor for a center feed strip dipole with length $\ell$ and width $\ell/100$. The Q factors are determined from the stored energies~\eqref{eq:WeF01} and~\eqref{eq:WmF01}, from differentiation of the input impedance~\eqref{eq:QZE} and~\eqref{eq:QZM}, and from Brune Synsesis~\eqref{eq:circuitQ}. a) electric and magnetic Q-factors from the subtracted far field $\QF$, the Brune synthesized circuit model $\QB$ (dotted), and differentiation of the impedance $\QZp$ (dashed dotted). b) difference between the computed Q-factors $\QF-\QZp$ and $\QF-\QB$.}
  \label{fig:DipoleQcomp1}
\end{figure}

Simple circuits models are accurate to model the input impedance over relatively narrow frequency bands. The accuracy is in general not sufficient over a wider frequency bands, see Fig.~\ref{fig:DipoleQcomp}. We use Brune synthesis~\cite{Brune1931} to construct more accurate wide band circuit models from the input impedance for the strip dipole. The resulting Q factor $\QB$ from the stored energy in the lumped elements~\eqref{eq:circuitQ} is depicted in Fig.~\ref{fig:DipoleQcomp1}a. It is seen that $\QF\approx\QB\approx\QZp$ for the considered range $\ell/\lambda<1.5$ or equivalently $ka<4.7$. The differences $\QF-\QB$ and $\QF-\QZp$ are also depicted in Fig.~\ref{fig:DipoleQcomp1}b. Here, we note that the differences are much less than $ka$ in contrast to the spherical mode case~\eqref{eq:QpE0}, see also~\eqref{eq:Qbound}.

\subsection{Loop antenna}
\begin{figure}[t]
\begin{center}
  \includegraphics[width=0.48\textwidth]{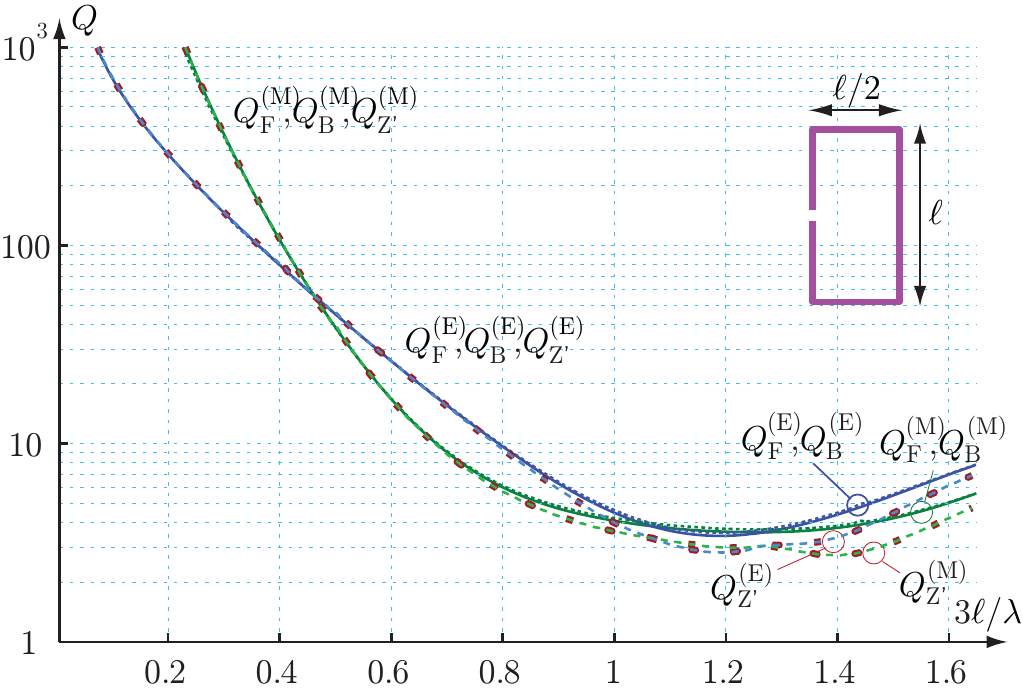}
	\hspace{1mm}
	\includegraphics[width=0.48\textwidth]{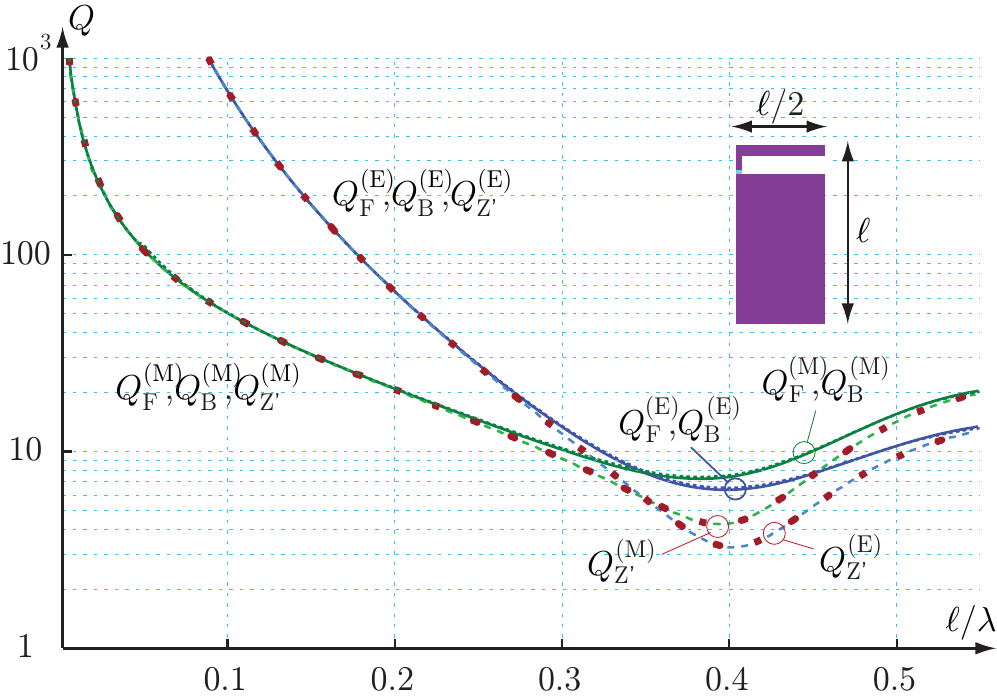}
\end{center}	
  \caption{Q factors for a rectangular loop antenna and an inverted L antenna with height $\ell$ and width $\ell/2$. The Q factors are determined from the stored energies~\eqref{eq:WeF01} and~\eqref{eq:WmF01}, differentiation of the impedance~\eqref{eq:QZE} and~\eqref{eq:QZM}, and the Brune synthesized lumped circuit~\eqref{eq:circuitQ}.}
  \label{fig:LoopQcomp}
	\label{fig:Mob1Qcomp}
\end{figure}

The computed stored electric and magnetic energies for a loop antenna are depicted in Fig.~\ref{fig:LoopQcomp}. The loop antenna is rectangular with height $\ell$, width $\ell/2$,  vanishing thickness, and is modeled as perfectly electric conducting (PEC). We see that the magnetic energy dominates for low frequencies and changes to dominantly electric energy at approximately $\ell\approx \lambda/6$ or equivalently $\mathcal{C}\approx\lambda/2$, where $\mathcal{C}=3\ell$ denotes the circumference of the loop. 

In Fig.~\ref{fig:LoopQcomp}, we see that the Q-factors determined from the stored energies~\eqref{eq:WeF01} and~\eqref{eq:WmF01} and from the Brune synthesized lumped circuit~\eqref{eq:circuitQ} agree very well. The Q factors also agree with the differentiation of the impedance for $Q\geq 10$. The difference increases between $\QZp$ and $\QF\approx\QB$ for lower Q values. This is consistent with the increasing difficulties to approximate the impedance with a single resonance model~\cite{Gustafsson+Nordebo2006b}. Here, it is also important to realize that the concept and usefulness of the Q-factor is increasingly questionable as $Q$ decreases towards unity, see also Sec.~\ref{S:ex_resonancecircuit}. 

\subsection{Inverted L antenna}
An inverted L antenna on a finite ground plane is considered to illustrate the usefulness of the stored energies for terminal antennas. The antenna has total length $\ell$ and width $\ell/2$, see Fig.~\ref{fig:Mob1Qcomp}. The electric and magnetic Q factors are depicted in  Fig.~\ref{fig:Mob1Qcomp}. Here, we see that $\QF$, $\QB$, and $\QZp$ agree well for $Q\geq 10$, that is for approximately $\ell\leq\lambda/3$ or below $1\unit{GHz}$ for $10\unit{cm}$ chassis. The results for $\QZp$ start to differ for larger structures, where 
\eg $\QFE\approx 5$ and $\QZpE\approx 2$ at $\ell/\lambda=0.4$ or $ka\approx 1.4$. For this levels of $\QZp$, the underlying single resonance model~\cite{Gustafsson+Nordebo2006b} is problematic and hence $\QZp$ reduces in accuracy. At the same time $Q$ is low enough to be considered less useful as a quantity to estimate the bandwidth, \eg $Q\approx 2$ corresponds to a half-power bandwidth of $100\%$.

\begin{figure}[t]
\begin{center}
  \includegraphics[width=0.48\textwidth]{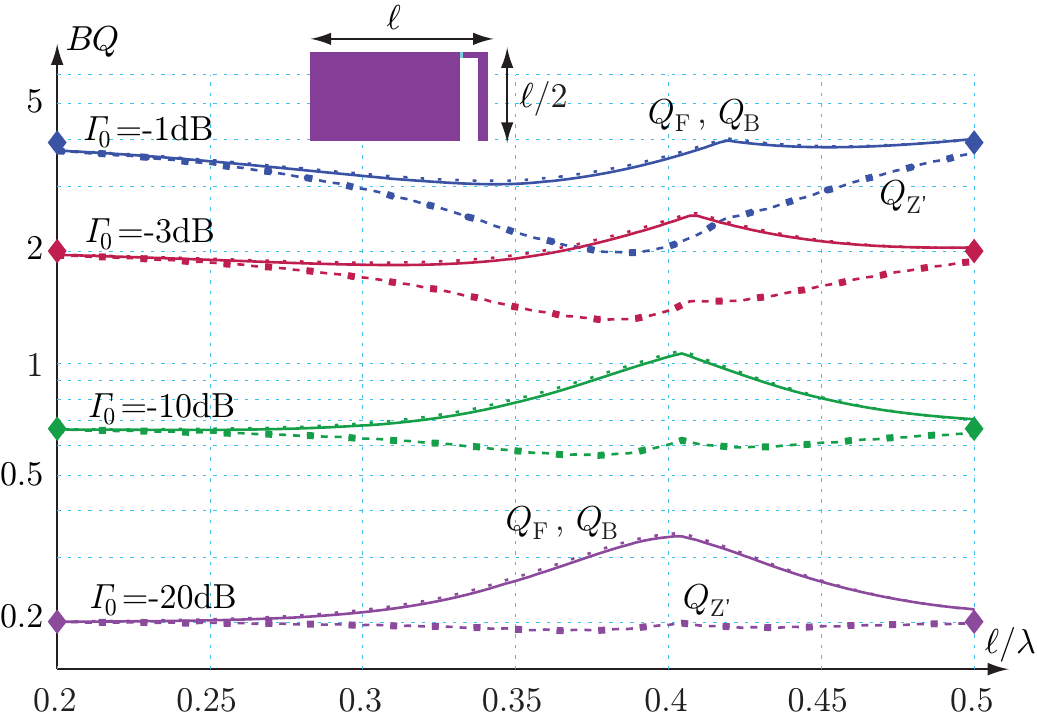}
	\hspace{1mm}
	\includegraphics[width=0.48\textwidth]{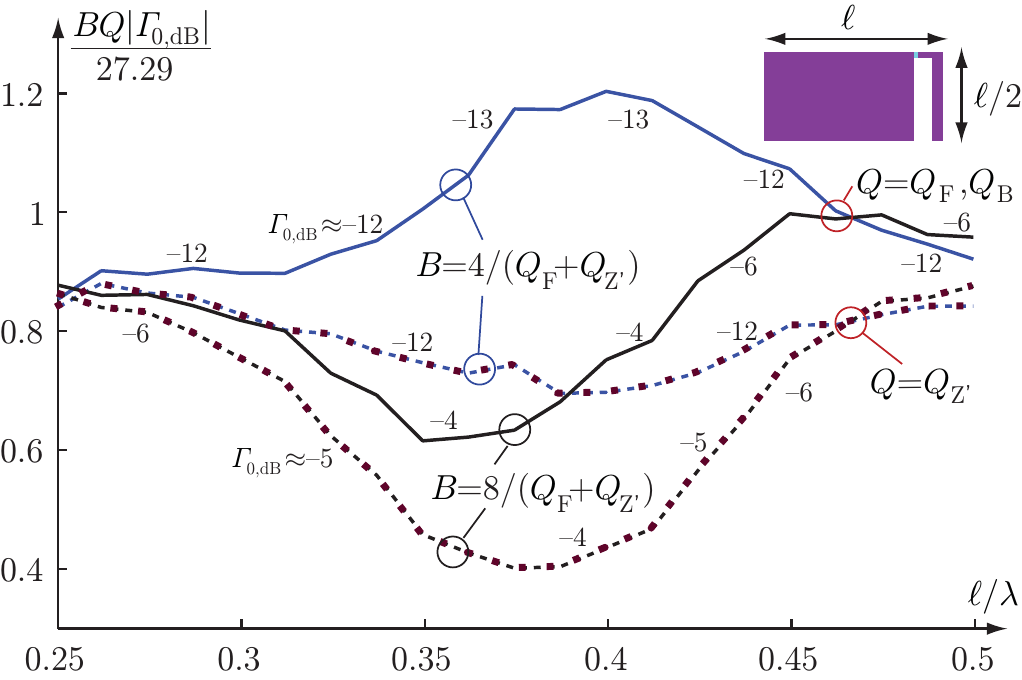}
\end{center}	
  \caption{Illustration of the Q factor fractional bandwidth product for the inverted L antenna using the Q from the far-field stored energy $\QF$, energy in the Brune synthesized lumped circuit model $\QB$, and differentiation of the input impedance $\QZp$. (left) The bandwidth is determined for the thresholds $\refl_0=-\{1,3,10,20\}\dB$ corresponding to the fractional bandwidths $B\approx\{48, 32, 14, 4.6\}\%$ at $\ell/\lambda=0.4$.
	(right) The real frequency technique~\cite{Carlin+Civalleri1998} is used to determine the matching threshold $\refl_0$ over the fractional bandwidths $B=\{2,4\}/\Qa$, where $\Qa$ denotes the mean $\Qa=(\QF+\QZp)/2$. The resulting $\refl_0$ are approximately $\refl_0\approx -12\dB$ for $B=2/\Qa$ and $\refl_0\approx -6\dB$ to $-4\dB$ for $B=2/\Qa$ as indicated in the graph.}
  \label{fig:Mob1_BQ}
\end{figure}

We use the fractional bandwidth for the antenna tuned to resonance with an inductor or capacitor to analyze the difference between the Q from the stored energy and differentiated impedance. The fractional bandwidth Q factor product, $BQ$, is given by~\eqref{eq:Q2Bandwidth} for simple RCL resonance circuits. The corresponding $BQ$ product for the inverted L antenna is depicted in Fig.~\ref{fig:Mob1_BQ} for the reflection coefficient thresholds $\refl_0=-\{1,3,10,20\}\dB$. It is seen that $BQ$ is close to the value given by~\eqref{eq:Q2Bandwidth}, $BQ\approx\{3.9,2.0,0.67,0.20\}$ as indicated by the rhombi, for $\ell/\lambda\leq 0.25$, where also $\QF\approx\QB\approx\QZp$. The $BQ$ product starts to deviates from~\eqref{eq:Q2Bandwidth} for shorter wavelengths except for the low reflection coefficient $\refl_0=-20\dB$ and $\QZp$ case. This is consistent with $\QZp$ being a local approximation of the Q-factor around the tuned resonance frequency and hence more accurate for relatively narrow fractional bandwidths $B\ll 1/Q$ or equivalently $\refl_0\ll 1$. The Q from $\QF$ and $\QB$ underestimate the fractional bandwidths for this case. The accuracy of $\QZp$ deteriorates as the threshold $\refl_0$ is relaxed and the relative bandwidth increases leading to increasingly difficulties to approximate the impedance with a resonance circuit. The results for the Q determined from the stored energy $\QF\approx\QB$ are on the contrary improving as the requirements on the matching are relaxed. This is consistent with $\QB$ being a global quantity determined from the input impedance over a large bandwidth. 

\section{Conclusions}\label{S:conclusions}
The analyzed expression~\eqref{eq:storedEF} for the stored energy defined by subtraction of the far-field energy density from the energy density is mainly motivated by the formulation of Collin \& Rothschild~\cite{Collin+Rothschild1964}, McLean~\cite{McLean1996}, Yaghjian \& Best~\cite{Yaghjian+Best2005} and the expressions by Vandenbosch in~\cite{Vandenbosch2010}.
We show that the stored energy~\eqref{eq:storedEF} is identical to the energy in~\cite{Vandenbosch2010} for many currents. However, some current densities have an additional coordinate dependent term. This term is very small for small antennas but it can contribute for larger structures, see also~\cite{Yaghjian+Best2005}. Here, it is also important to realize that the classical definition~\cite{Collin+Rothschild1964} with the subtracted power flow~\eqref{eq:storedEP} is inherently coordinate dependent. The identification of the energy expressions in~\cite{Vandenbosch2010} with~\eqref{eq:storedEF} offers a simple interpretation of the observed cases with a negative stored energy~\cite{Gustafsson+etal2012a}. The analysis also suggests that the resulting Q factor has an uncertainty of the order $ka$. This is consistent with the use of the results for small (sub wavelength) antennas~\cite{Gustafsson+etal2012a,Gustafsson+Nordebo2013}, where $ka$ is small and Q large. 

The energy expressions proposed by Vandenbosch in~\cite{Vandenbosch2010}  are very well suited for optimization formulations as they are simple quadratic forms of the current density. The quadratic form is very practical as it allows for various optimization formulations such as Lagrangian~\cite{Gustafsson+etal2012a} and convex optimization~\cite{Gustafsson+Nordebo2013} and has already led to many new antenna results. Their resemblance of the electric field integral equation (EFIE) makes the numerical implementation very simple. Analytic solutions for spherical structures show that the Q in~\cite{Vandenbosch2010} and~\cite{Hansen+Collin2009} differ by $ka$ and this is interpreted as the far-field in the interior of the sphere as seen from~\eqref{eq:storedEF} and~\eqref{eq:storedEP}. The new formulation also produce simplified expressions~\eqref{eq:QpE0} the unifies the TE and TM cases for the Q factor of current densities on spherical shells~\cite{Hansen+Collin2009}. 

Numerical results for dipole, loop, and inverted L antennas are used to illustrate the accuracy of the energy expressions. The Q factors from the stored energy in the fields, $\QF$, from the stored energy in Brune synthesized circuit models, $\QB$, and from differentiation of the input impedance, $\QZp$ are compared. It is observed that $\QF\approx\QB$ for the considered cases. The good agreement is based on Brune circuits synthesized from the input impedance over a wide frequency range and it is observed that $\QB$ can be lower if a narrow range of frequencies is used. We also observe that $\QF\approx\QB\approx\QZp$ in the regions where Q is $\QF\geq 10$ and that $\QZp\leq \QF\approx\QB$ otherwise. This is consistent with the shunt and series resonance circuit that can have $\QZp=0$, see~\eqref{eq:RCLflatmatch} and~\cite{Gustafsson+Nordebo2006b}.

The results for the spherical modes and the antennas are very interesting for the interpretation of the Q factor, $\QF$, defined by subtraction of the far field~\eqref{eq:storedEF}. The analytic results for the spherical modes show that the subtracted far-field inside the sphere defines a stored energy that can be negative, see also~\cite{Gustafsson+etal2012a}, and a $\QF$ that differ from the classical definition~\eqref{eq:storedEP} by $ka$, \ie $\QP=\QF+ka$, where $\QP$ denotes the Q defined using~\eqref{eq:storedEP}. Contrary, the numerical results for the simulated antennas show that $\QF$ agree with the Q from the stored energy in the equivalent circuit synthesized from the input impedance using Brune synthesis, \ie $\QF\approx \QB$. This suggests that the Q from the subtracted far-field is \textit{correct} for the tested antennas but not for the spherical modes.
This is consistent with the interpretation~\eqref{eq:Qbound} that shows that $\QF$ has an uncertainty of the order $ka$.

\section{Acknowledgments}
This work was supported by the Swedish Research Council (VR), the Crafoord Foundation, and the Swedish Governmental Agency for Innovation Systems (VINNOVA) within the VINN Excellence Center CHASE. 

Prof. Edwin Marengo at Northeastern University, MA, USA is also acknowledged for valuable discussions and his hospitality during a visit of MG to Boston in August 2012.

\appendix

\section{Green's function identities}\label{sec:GFI}
Multiply the Helmholtz Green's function for $G_1$: $(\nabla^2+ k^2)G_1=-\delta(\rv - \rv_1)$ with $G_2^*$, and similarly for $G_2^*$. Adding the results together with a standard vector calculus identity gives $2(\nabla G_1\cdot \nabla G_2^* - k^2 G_1G_2^*) = G_1\delta_2 + G_2^* \delta _1 + \nabla^2(G_1G_2^*)$, where $\delta_n=\delta(\rv-\rv_n)$ with $n=1,2$ denotes the Dirac delta distribution. Integration yields the identity~\cite{Vandenbosch2010}
\begin{equation}\label{c1}
	\int_{\Rr^3}\nabla G(\rv-\rv_1)\cdot\nabla G^{\ast}(\rv-\rv_2)
	-k^2G(\rv-\rv_1)G^{\ast}(\rv-\rv_2)\diffV
	=\frac{\cos(k|\rv_1-\rv_2|)}{4\pi|\rv_1-\rv_2|},
\end{equation}
where we used Gauss's theorem together with the observation that 
\begin{equation}
	\nabla(G_1G_2^*)\rightarrow \frac{-\rvh\eu^{\iu k \rvh\cdot (\rv_2-\rv_1)}}{8\pi^2 r^3}
\end{equation}
 for large enough radius. 

The $k$-derivative of the Helmholtz Green's equation for $G_1$ is $(\nabla^2 + k^2)\partial_k G_1 + 2kG_1 = 0$. Similarly to the derivation of \eqref{c1} we multiply with $G_2^*$, and repeat the procedure with the $k$-derivative of $G_2^*$. Adding the result and applying vector calculus identities to move $\nabla^2$ away from the $k$-derivative results in the identity
\begin{equation}
4k G_1G_2^* = \delta_2\partial_k G_1 + \delta_1\partial_k G_2^* - \nabla\cdot\qv,
\end{equation}
where
\begin{multline}
\rvh\cdot\qv= \rvh\cdot\big(G_1\nabla\partial_k G_2 - (\partial_k G_2^*)\nabla G_1 + G_2^*\nabla\partial_k G_1 
- (\partial_k G_1)\nabla G_2^*\big)
\\
\rightarrow
\frac{-k}{8\pi^2r} \Big[2+\frac{1}{r}\big(\rvh\cdot(\rv_1+\rv_2)
+\iu (|\rvh\times\rv_1|^2-|\rvh\times\rv_2|^2)\big)
+\Ordo(\frac{1}{r^2})\Big]\eu^{-\iu k \rvh\cdot(\rv_1-\rv_2)}
\end{multline}
for large enough radius. 
Collecting term of decay rate $r^{-1}$ on the left-hand side and the remaining terms on the right-hand side. Integration over a large sphere, together with Gauss's theorem and elementary integrals results in 
\begin{multline}\label{eq:GreenFId1}
	\int_{\Rr^3} G(\rv-\rv_1)G^{\ast}(\rv-\rv_2)-\frac{\eu^{-\iu k(\rv_1-\rv_2)\cdot\rvh}}{16\pi^2r^2}\diffV\\
	=-\frac{\sin(kr_{12})}{8\pi k}  +\iu \frac{r_1^2-r_2^2}{8\pi k^2 r_{12}^3} (\sin (kr_{12})-kr_{12}\cos(kr_{12}))\\
	=-\frac{\sin(kr_{12})}{8\pi k}
	+\iu\frac{(\rv_1+\rv_2)\cdot(\rv_1-\rv_2)}{8\pi r_{12}}\jop_1(kr_{12})\\
	=-\frac{\sin(kr_{12})}{8\pi k}-\iu\frac{(\rv_1+\rv_2)}{k}\cdot\nabla_1\frac{\sin(kr_{12})}{8\pi kr_{12}}.
\end{multline}
Here $\jop_1(z)=(\sin(z)-z\cos(z))/z^2$ and $r_{12}=|\rv_1-\rv_2|$. Note that~\eqref{eq:GreenFId1} generalizes the result in~\cite{Vandenbosch2010} to the case $\rv_1+\rv_2\neq\vec{0}$ and shows that the integral depends of the coordinate system. The result also shows that it is necessary to specify how the integration over $\R^3$ is performed, \ie here as the limit $\Rr^3=\{\rv: \lim_{r_0\to\infty}|\rv|<r_0\}$.

\section{Electric surface currents on a sphere}\label{S:Sphere} 
The two formulations~\eqref{eq:storedEP} and~\eqref{eq:storedEF} for the stored energy can be compared for electric surface currents on spherical shells. This is the case analyzed by Thal~\cite{Thal2006} and Hansen \& Collin~\cite{Hansen+Collin2009}, see also~\cite{Hansen+etal2012} for the case with electric and magnetic surface currents.
We expand the surface current on a sphere with radius $a$ in vector spherical harmonics $\Yvop$, see App.~\ref{S:sphwaves}. For simplicity, consider the surface current $\Jv(\rv)=J_0\Yvop_{\tau\sigma ml}(\rvh)\delta(r-a)$ that induces the electric and magnetic fields 
\begin{equation}\label{eq:spherefields}
	\Ev(\rv) = \iu\eta_0\tilde{J}_0 
		\frac{\uop^{(p)}_{\tau\sigma ml}(k\rv)}{\Rop^{(p)}_{\tau l}(ka)}  
		\text{ and }
	\Hv(\rv) = \tilde{J}_0 
		\frac{\uop^{(p)}_{\dtau\sigma ml}(k\rv)}{\Rop^{(p)}_{\tau l}(ka)},  
\end{equation}
where $p=1$ for $r<a$ and $p=3$ for $r>a$, $\uop^{(p)}_{\tau \sigma ml}$ is the spherical vector waves, and $\Rop^{(p)}_{\tau l}$ the radial functions in Hansen~\cite{Hansen1988}, defined as 
\begin{equation}\label{eq:radialfunctions}
	\Rop_{\tau l}^{(p)}(\kappa)
	=\begin{cases}
		z_l^{(p)}(\kappa) & \tau=1\\
		\displaystyle{\frac{1}{\kappa}\partder{(\kappa z_l^{(p)}(\kappa))}{\kappa}}
		& \tau=2,
	\end{cases}
\end{equation}
where $z_l^{(1)}=\jop_l$ are Bessel functions, $z_l^{(2)}=\nop_l$ Neumann functions, $z_l^{(3)}=\hop^{(1)}_l$ Hankel functions~\cite{Hansen1988}, and $\kappa=ka$. We note that the derivatives of $\Rop_{\tau l}^{(p)}(\kappa)$ are easily expressed in $z^{(p)}$, see App.~\ref{S:sphwaves}. Here, $\tau=1$ is transverse electric (TE) and $\tau=2$ transverse magnetic (TM) waves. Moreover, the dual index $\dtau$ is $\dtau=2$ if $\tau=1$ and $\dtau=1$ if $\tau=2$. The current in~\eqref{eq:spherefields} is rescaled as $\tilde{J}_0=J_0\Rop^{\mrm{(1)}}_{\tau l}(ka)\Rop^{\mrm{(3)}}_{\tau l}(ka)$
 and below we let $J_0$ be real valued to simplify the notation. 
We also note that the coordinate dependent term~\eqref{eq:WemF2} vanishes for single spherical modes.

\subsection{Far-field type stored energy for the TE case $W_{\mrm{F}}$}
We start with the transverse electric (TE) case $\tau=1$, \ie $\Jv(\rv)=\Yvop_{1\sigma ml}(\rvh)\delta(r-a)$ that is divergence free, $\nabla\cdot\Jv=0$. The integrals in~\eqref{eq:WeF} are evaluated analytical by expanding the Green's functions in~\eqref{eq:ZeJ}, \eqref{eq:ZmJ}, and~\eqref{eq:ZemJ} in spherical modes, see App.~\ref{S:sphwaves}. Using $\nabla\cdot\Yvop_{1\sigma ml}=0$, we get
$\langle \Jv,\Lop_\mrm{e}\Jv\rangle=0$ for~\eqref{eq:ZeJ} and hence the first part of the stored electric energy $W^{\mrm{(E)}}_{\mrm{F_0}}=0$.
The expansion of the full Green's dyadic, $\mat{G}=G\Id$,~\eqref{eq:GreenFDyadicExp} gives
\begin{multline}
	\frac{1}{\iu k J_0^2}\langle \Jv,\Lop_\mrm{m}\Jv\rangle
	=\int_{V}\!\int_{V}\Yvop_{1\sigma ml}(\rvh_1)\delta(r_1-a)\cdot \mat{G}(\rv_1-\rv_2)\cdot\Yvop_{1\sigma ml}(\rvh_2)\delta(r_2-a)\diffVa\diffVb\\
	=
	a^4\int_{\Omega}\!\int_{\Omega} \Yvop_{1\sigma ml}(\rvh_1)\cdot\mat{G}(\rv_1-\rv_2)\cdot\Yvop_{1\sigma ml}(\rvh_2)\diff\Omega_1\diff\Omega_2
	=\iu a^4 k\Rop_{1l}^{(3)}(\kappa)\Rop_{1l}^{(1)}(\kappa)
\end{multline}	
for the terms in~\eqref{eq:ZmJ} to get the first part of the stored magnetic energy from~\eqref{eq:ZtoWM} as $4\omega\eta_0^{-1} W_{\mrm{F_0}}^{\mrm{(M)}}= -a^2\kappa^2J_0^{2}\Rop_{1l}^{(2)}\Rop_{1l}^{(1)}$. The radiated power follow from~\eqref{eq:ZtoP} $2\eta_0^{-1}P_{\mrm{r}}=-\Re\langle\Jv,\Lop_\mrm{m}\Jv \rangle
=a^2\kappa^2 J_0^{2} (\Rop_{1l}^{(1)})^2$.
The corresponding expansion of the frequency derivative of the Green's function~\eqref{eq:GreenFDyadicExp} is used for the terms related to~\eqref{eq:ZemJ}
\begin{multline}\label{eq:sphQTMem}
	\frac{-2}{\iu k^2a^4 J_0^{2}}\langle \Jv,\Lop_\mrm{em}\Jv\rangle
	=\int_{\Omega}\!\int_{\Omega} \Yvop_{1\sigma ml}(\rvh_1)\cdot\partder{\mat{G}(\rv_1-\rv_2)}{k}\cdot\Yvop_{1\sigma ml}(\rvh_2)\diff\Omega_1\diff\Omega_2\\	=\iu\partder{}{\kappa}\left(\kappa\Rop_{1l}^{(3)}(\kappa)\Rop_{1l}^{(1)}(\kappa)\right)
	=\iu\big(\kappa\Rop_{1l}^{(3)}(\kappa)\Rop_{1l}^{(1)}(\kappa)\big)'\\
	=\iu(\Rop_{1l}^{(3)}\Rop_{1l}^{(1)}+\kappa \Rop_{1l}^{(3)}{}'\Rop_{1l}^{(1)}
	+\kappa \Rop_{1l}^{(3)}\Rop_{1l}^{(1)}{}'),	
\end{multline}
where ${}'$ denotes differentiation with respect to $\kappa$, giving $4\omega\eta_0^{-1}W_{\mrm{F_1}}=-\frac{a^2\kappa^2}{2}J_0^{2}(\kappa\Rop_{1l}^{(2)}\Rop_{1l}^{(1)})'$.

Collecting the terms gives the electric and magnetic Q-factors as
\begin{equation}\label{eq:Qi1E}
	Q^{\mrm{(E)}}_{1l,\mrm{F}}(\kappa) 
	=\frac{2\omega W_{\mrm{F}}^{\mrm{(E)}}(\kappa)}{P_{\mrm{r}}(\kappa)}	=-\frac{\big(\kappa\Rop_{1l}^{(1)}(\kappa)\Rop_{1l}^{(2)}(\kappa)\big)'}{2 (\Rop_{1 l}^{(1)}(\kappa))^2}
\end{equation}
and
\begin{equation}\label{eq:Qi1M}
	Q^{\mrm{(M)}}_{1l,\mrm{F}}(\kappa) 
	=\frac{2\omega W_{\mrm{F}}^{\mrm{(M)}}(\kappa)}{P_{\mrm{r}}(\kappa)}
	=Q^{\mrm{(E)}}_{1l,\mrm{F}}(\kappa) 
	-\frac{\Rop_{1 l}^{(2)}(\kappa)}{\Rop_{1 l}^{(1)}(\kappa)},
\end{equation}
respectively.
We note that $\Rop_{1 l}^{(1)}=\jop_l$ and $\Rop_{1 l}^{(2)}=\nop_l$ can be used to rewrite the Q-factors, however the form with the radial functions simplifies the comparison with the TM case below. The differentiated terms are also easily evaluated using~\eqref{eq:sphQTMem} and~\eqref{eq:radialfuncprime}.

\subsection{Far-field type stored energy for the TM case $W_{\mrm{F}}$}
The transverse magnetic (TM) case is given by $\tau=2$ and generated by the current density $\Jv(\rv)=J_0\Yvop_{2\sigma ml}(\rvh)\delta(r-a)$ that has the divergence $\nabla\cdot\Yvop_{2\sigma ml}=-\sqrt{l(l+1)}\Yop_{\sigma ml}/r$. With the expansion of the Green's function~\eqref{eq:GreenfuncExp} we get the part related to the charge density~\eqref{eq:ZeJ}
\begin{multline}
	\frac{-\iu k}{a^{4}J_0^{2}}\langle \Jv,\Lop_\mrm{e}\Jv\rangle
	=\int_{\Omega}\!\int_{\Omega}\nabla_1\cdot\Yvop_{2\sigma ml}(\rvh_1)G(\rv_1-\rv_2)\nabla_2\cdot\Yvop_{2\sigma ml}(\rvh_2)
	\diff\Omega_1\diff\Omega_2\\
	=\frac{\iu k l(l+1)}{a^2}\jop_l(\kappa)\hop_l^{(1)}(\kappa)
\end{multline}
and the full Green's Dyadic expansion~\eqref{eq:GreenFDyadicExp} gives
\begin{multline}
\frac{1}{\iu k a^{4}J_0^{2}}\langle \Jv,\Lop_\mrm{m}\Jv\rangle
	=\int_{\Omega}\!\int_{\Omega}
	\Yvop_{2\sigma ml}(\rvh_1)\cdot\mat{G}(\rv_1-\rv_2)\cdot\Yvop_{2\sigma ml}(\rvh_2)
	\diff\Omega_1\diff\Omega_2
	\\
	=\iu k\left(\Rop_{2 l}^{(1)}(\kappa)\Rop_{2 l}^{(3)}(\kappa)
	+l(l+1)\frac{\hop_l^{(1)}(\kappa)\jop_l(\kappa)}{\kappa^2}\right)
\end{multline}
for the part related to the current density~\eqref{eq:ZmJ}.
The expansions of the frequency derivatives of the Green's function~\eqref{eq:GreenfuncExp} and Green's Dyadic~\eqref{eq:GreenFDyadicExp} give
\begin{multline}
	\Re\int_{\Omega}\!\!\int_{\Omega}
	\Yvop_{2\sigma ml}(\rvh_1)\cdot\frac{\partial\mat{G}(\rv_1-\rv_2)}{\partial k}\cdot\Yvop_{2\sigma ml}(\rvh_2)\\
	-\nabla_1\cdot\Yvop_{2\sigma ml}(\rvh_1)\frac{\partial\mat{G}(\rv_1-\rv_2)}{k^2\partial k}\nabla_2\cdot\Yvop_{2\sigma ml}(\rvh_2)
	\diff\Omega_1\diff\Omega_2\\
	=2l(l+1)\nop_l(\kappa)\jop_1(\kappa)
	-\kappa^2(\kappa \Rop_{2 l}^{(1)}(\kappa)\Rop_{2 l}^{(2)}(\kappa))'
\end{multline}
for the part related to~\eqref{eq:ZemJ}.

Collecting the terms gives that the normalized radiated power is\\ $2\eta_0^{-1}P_{\mrm{r}}/J_0^2=\Re\langle\Jv,(\Lop_\mrm{e}-\Lop_\mrm{m})\Jv \rangle/J_0^2=a^3\kappa (\Rop_{2l}^{(1)})^2$.
The electric and magnetic Q factors are finally determined to
\begin{equation}\label{eq:Qi2E}
	Q^{\mrm{(E)}}_{2l,\mrm{F}}(\kappa)  
	=-\frac{\big(\kappa\Rop_{2l}^{(1)}(\kappa)\Rop_{2l}^{(2)}(\kappa)\big)'}{2 (\Rop_{2 l}^{(1)}(\kappa))^2}
\end{equation}
and
\begin{equation}\label{eq:Qi2M}
	Q^{\mrm{(M)}}_{2l,\mrm{F}} 
	=Q^{\mrm{(E)}}_{2l,\mrm{F}}(\kappa) 
	-\frac{\Rop_{2l}^{(2)}(\kappa)}{\Rop_{2l}^{(1)}(\kappa)},
\end{equation}
respectively. We note that the expressions for the TE case in~\eqref{eq:Qi1E} and~\eqref{eq:Qi1M} and TM case in~\eqref{eq:Qi2E} and~\eqref{eq:Qi2M} are written in identical forms by using the radial functions~\eqref{eq:radialfunctions}.

\subsection{Power flow stored energy $W_{\mrm{P}}$}
The stored electric energy with the subtracted power flow~\eqref{eq:storedEP} is analyzed by Hansen \& Collin~\cite{Hansen+Collin2009}, see also~Thal~\cite{Thal2006}. The integral~\eqref{eq:storedEP} is decomposed into integration of the exterior and interior regions where we have outgoing waves, $\uop_{\tau\sigma ml}^{(3)}$, and regular waves, $\uop_{\tau\sigma ml}^{(1)}$, respectively in~\eqref{eq:spherefields}. The exterior part was already analyzed by Collin \& Rothschild~\cite{Collin+Rothschild1964}. 
The subtracted power flow in~\eqref{eq:storedEP} of the fields~\eqref{eq:spherefields} has the radial dependence
\begin{equation}
	P_{\mrm{r}}=
	\frac{1}{2}
	\Re\int_{\Omega}\Ev(\rv)\times\Hv^{\ast}(\rv)\cdot\rvh r^2\diff\Omega
	=\frac{\tilde{J}_0^2\eta_0}{2|\Rop_{\tau l}^{(3)}(\kappa)|^2}
\end{equation}
in the exterior region $r\geq a$ and vanishes in the interior region $r<a$. 
As the spherical vector waves are orthogonal over the unit sphere they can be analyzed separately. Their integrals are divided into its angular and radial parts. 
To simplify the notation, we introduce the normalized energies $w_{\tau l}^{\mrm{(e)}}$ and $w_{\tau l}^{\mrm{(i)}}$ outside and inside the sphere, respectively. They are given by, see App.~\ref{S:SVWvolymeint} for details  
\begin{multline}
	w_{1l}^{\mrm{(e)}}
	=\int_{\kappa}^{\infty}\int_{\Omega} |\uop^{(3)}_{1\sigma ml}(k\rv)|^2 
	 k^2r^2\diff\Omega -1 \diff kr\\
	=\kappa-\frac{\kappa^3}{2}(|\hop^{\mrm{(1)}}_l(\kappa)|^2 - \Re\{\hop^{\mrm{(1)}}_{l+1}(\kappa)\hop_{l-1}^{\mrm{(2)}}(\kappa)\})
\end{multline}
for $\tau=1$ and for the $\tau=2$ modes
\begin{equation}
	w_{2l}^{\mrm{(e)}}
	=\int_{\kappa}^{\infty}\int_{\Omega} |\uop^{(3)}_{2\sigma ml}(k\rv)|^2 
	 k^2r^2\diff\Omega -1 \diff kr
	=-\Re\{\kappa\hop_l^{\mrm{(2)}}(\kappa)(\kappa\hop^{\mrm{(1)}}_l(\kappa))'\}
	+w_{1l}^{\mrm{(e)}}.
\end{equation}

The corresponding normalized energy in the interior of the sphere is given by the integrals
\begin{multline}
	w_{1l}^{\mrm{(i)}}
	=\int_0^{\kappa}\int_{\Omega} |\uop^{(1)}_{1 \sigma ml}(k\rv)|^2 
	 k^2r^2\diff\Omega  \diff kr
	=\int_0^{\kappa} x^2|\jop_l(x)|^2 \diff x\\
	=\frac{\kappa^3}{2}\left(\jop_l^2(\kappa)-\jop_{l-1}(\kappa)\jop_{l+1}(\kappa)\right)
\end{multline}
and
\begin{equation}
	w_{2l}^{\mrm{(i)}}
		=\int_0^{\kappa}\int_{\Omega} |\uop^{(1)}_{2\sigma ml}(k\rv)|^2 
	 k^2r^2\diff\Omega  \diff kr
	=-\Re\{\kappa\jop_l(\kappa)(\kappa\jop_l(\kappa))'\}
	+w_{1l}^{\mrm{(i)}}.
\end{equation}

We have the electric and magnetic Q factors
\begin{equation}
	Q^{\mrm{(E)}}_{\tau l,\mrm{P}}
	=|\Rop^{(3)}_{\tau l}(\kappa)|^2
	\left(\frac{w^{\mrm{(e)}}_{\tau l}(\kappa)}{|\Rop^{(3)}_{\tau l}(\kappa)|^2}
	+\frac{w^{\mrm{(i)}}_{\tau l}(\kappa)}{|\Rop^{(1)}_{\tau l}(\kappa)|^2}\right)
\end{equation}
and
\begin{equation}
	Q^{\mrm{(M)}}_{\tau l,\mrm{P}}
	=|\Rop^{(3)}_{\tau l}(\kappa)|^2
	\left(\frac{w^{\mrm{(e)}}_{\dtau l}(\kappa)}{|\Rop^{(3)}_{\tau l}(\kappa)|^2}
	+\frac{w^{\mrm{(i)}}_{\dtau l}(\kappa)}{|\Rop^{(1)}_{\tau l}(\kappa)|^2}\right).
\end{equation}
After extensive simplifications we can rewrite them as
\begin{equation}\label{eq:QpE}
	Q^{\mrm{(E,M)}}_{\mrm{\tau,P}}(\kappa)
	=\frac{2\omega W^{\mrm{(E,M)}}_{\mrm{P}}(\kappa)}{P_{\mrm{r}}(\kappa)}
	=\kappa+Q^{\mrm{(E,M)}}_{\mrm{\tau,F}}(\kappa),
\end{equation}
where $Q^{\mrm{(E,M)}}_{\mrm{\tau,F}}$ denotes the electric and magnetic far-field type Q factors in~\eqref{eq:Qi1E},~\eqref{eq:Qi1M},~\eqref{eq:Qi2E}, and~\eqref{eq:Qi2M}. Note that the subscript EM is used to denote E and M in~\eqref{eq:QpE}.
The difference $\kappa=ka$ is consistent with the interpretation of a standing wave in the interior of the sphere, \cf~\eqref{eq:Qbound}. Moreover, the expressions~\eqref{eq:Qi2E} and~\eqref{eq:Qi2M} unifies the TE and TM cases and offer an alternative to the expressions in~\cite{Hansen+Collin2009}, here we also note a misprint in (6) in~\cite{Hansen+Collin2009}.

\subsection{Spherical waves}\label{S:sphwaves}
The radiated electromagnetic field is expanded in spherical vector waves or modes~\cite{Hansen1988}:
\begin{equation}\label{eq:sphericalvw}
\begin{cases}
	\uop^{(p)}_{1\sigma ml}(k\rv)= \Rop_{1l}^{(p)}(kr)\Yvop_{1\sigma ml}(\rvh)\\
	\uop^{(p)}_{2\sigma ml}(k\rv)= \Rop_{21}^{(p)}(kr)\Yvop_{2\sigma ml}(\rvh)
	+\widetilde{\Rop}(kr)\Yop_{\sigma ml}(\rvh)\rvh\\
	\uop^{(p)}_{3\sigma ml}(k\rv)= z_l^{(p)\prime}(kr)\Yop_{\sigma ml}(\rvh)\rvh
	+\widetilde{\Rop}(kr)\Yvop_{2\sigma ml}(\rvh)
	\end{cases}
\end{equation}
where $\rv$ is the spatial coordinate, $\rvh=\rv/r$, $r=|\rv|$, $k$ the wavenumber,\\ $\widetilde{\Rop}(\kappa)=\sqrt{l(l+1)}z_l^{(p)}(\kappa)/\kappa$, and $\Rop_l^{(p)}(kr)$ are the radial function of order $l$:
\begin{equation}
	\Rop_{\tau l}^{(p)}(\kappa)
	=\begin{cases}
		z_l^{(p)}(\kappa) & \tau=1\\
		\displaystyle{\frac{1}{\kappa}\partder{(\kappa z_l^{(p)}(\kappa))}{\kappa}} & \tau=2.
	\end{cases}
\end{equation}
For regular waves ($p=1$) $z_l^{(1)}=\jop_l$ is a spherical Bessel function, irregular waves ($p=2$) $z_l^{(2)}=\nop_l$ is a spherical Neumann function, and outgoing waves ($p=3$) $z_l^{(3)}=\hop^{(1)}_l$ is an outgoing spherical Hankel function. The indices are $\sigma=\{\mrm{e,o}\}$, $m=0,\ldots,l$, $l=1,\ldots$~\cite{Harrington1961,Bostrom+Kristensson+Strom1991}.
In addition, 
$\Yvop_{\tau\sigma ml}(\rvh)$ denotes the vector spherical harmonics defined as 
\begin{equation}\label{eq:svh}
	\Yvop_{1\sigma ml}(\rvh)
	= \frac{1}{\sqrt{l(l+1)}}\nabla\times\big(\rv\Yop_{\sigma ml}(\rvh)\big)
\end{equation}
and $\Yvop_{2\sigma ml}(\rvh)=\rvh\times\Yvop_{1\sigma ml}(\rvh)$
where $\Yop_{\sigma ml}$ denotes the ordinary spherical harmonics~\cite{Bostrom+Kristensson+Strom1991}. There are a few alternative definitions of the spherical vector waves in the literature~\cite{Harrington1961,Hansen1988,Bostrom+Kristensson+Strom1991}. Here, we follow~\cite{Harrington1961} and use $\cos m\phi$ and $\sin m\phi$ as basis functions in the azimuthal coordinate. This choice is motivated by the interpretation of the fields related to the first 6 modes as the fields from different Hertzian dipoles. The modes labeled by $\tau=1$ are TE modes (or magnetic $2^{l}$-poles) while those labeled by $\tau=2$ correspond to TM modes (or electric $2^{l}$-poles). 
We note that the derivatives of $\Rop_{\tau n}^{(p)}(\kappa)$ are easily expressed in spherical Bessel and Hankel functions, \ie
\begin{equation}\label{eq:radialfuncprime}
	\partder{\Rop_{\tau l}^{(p)}}{\kappa}
	=\begin{cases}
		\displaystyle{\partder{}{\kappa}z_l^{(p)}} & \tau=1\\
		\displaystyle{
		\frac{-\Rop_{\tau l}^{(p)}}{\kappa}+\frac{l(l+1)-\kappa^2}{\kappa^2}z_l^{(p)}} & \tau=2.
	\end{cases}
\end{equation}
The Green functions are expanded in spherical waves to analyze spherical geometries.
The scalar Green's function has the expansion~\cite{Bostrom+Kristensson+Strom1991}
\begin{equation}\label{eq:GreenfuncExp}
	G(\rv_1-\rv_2)=\frac{\eu^{\iu k|\rv_1-\rv_2|}}{4\pi|\rv_1-\rv_2|}
	=\iu k\sum_{\sigma ml}
	\jop_{l}(kr_<)\hop_l^{(1)}(kr_>)\Yop_{\sigma ml}(\rvh_1)\Yop_{\sigma ml}(\rvh_2),
\end{equation}
where $r_<=\min\{|\rv_1|,|\rv_2|\}$ and $r_>=\max\{|\rv_1|,|\rv_2|\}$, and $\Yop_{\sigma ml}$ denotes the spherical harmonics. 
In addition, the full Green's dyadic, $\Gm=\Id G$, can be expanded as~\cite{Bostrom+Kristensson+Strom1991}
\begin{equation}\label{eq:GreenFDyadicExp}
	\Gm(\rv_1-\rv_2)=\iu k\sum_{\tau\sigma m l} \uop^{\mrm{(1)}}_{\tau\sigma m l}(k\rv_<)\uop^{\mrm{(3)}}_{\tau\sigma m l}(k\rv_>),
\end{equation}
where $\tau=1,2,3$.
We also use the frequency derivatives of the Green's function and the Green's dyadic expansions.

\subsection{Volume integrals}\label{S:SVWvolymeint}
The volume integrals of the spherical vector waves are given by integrals of spherical Hankel functions as evaluated here.
We have
\begin{equation}
	\int x^2 z_p^2(x)\diff x
	=\frac{x^3}{2}\left(z_p^2(x)-z_{p-1}(x)z_{p+1}(x)\right).
\end{equation}
For the spherical Hankel function $z_p=\hop^{(1)}_p=\jop_p+\iu\nop_p$ we have
\begin{equation}\label{eq:hankelint1}
	\int x^2 |\hop^{(1)}_p(x)|^2\diff x\\
	=\frac{x^3}{2}\left(|\hop^{(1)}_p(x)|^2-\Re\{\hop^{(1)}_{p-1}(x)\hop^{(1)\ast}_{p+1}(x)\}\right).
\end{equation}
To evaluate the stored reactive energy outside a sphere, we need the result
\begin{equation}
	\int_a^{\infty} \big( x^2 |\hop^{(1)}_p(x)|^2-1 \big) \diff x
	=a-\frac{a^3}{2}
	\left(|\hop^{(1)}_p(a)|^2-\Re\{\hop^{(1)}_{p-1}(a)\hop^{(1)\ast}_{p+1}(a)\}\right).
\end{equation}
The corresponding internal energy is
\begin{equation}
	\int_0^{a} x^2 |\jop^{(2)}_p(x)|^2\diff x
	=\frac{a^3}{2}\left(\jop_p^2(a)-\jop_{p-1}(a)\jop_{p+1}(a)\right).
\end{equation}
We also have for $\tau=1$ 
\begin{equation}
	\int_{[a,b]\times\Omega}|\uop_{1\sigma ml}(k\rv)|^2\diffV
	=\int_a^b|\hop^{(1)}_l(kr)|^2 r^2\diff r
\end{equation}
For $\tau=2$, we use
\begin{multline}
	k|\uop_{2\sigma ml}|^2
	=k\uop_{2\sigma ml}\cdot\uop_{2\sigma ml}^{\ast}
	=\nabla\times\uop_{1\sigma ml}\cdot\uop_{2\sigma ml}^{\ast}\\
	=\nabla\cdot(\uop_{1\sigma ml}\times\uop_{2\sigma ml}^{\ast})
	+\uop_{1\sigma ml}\cdot\nabla\times\uop_{2\sigma ml}^{\ast}
=\nabla\cdot(\uop_{1\sigma ml}\times\uop_{2\sigma ml}^{\ast})
	+k|\uop_{1\sigma ml}|^2
\end{multline}
and hence
\begin{equation}
	\int_{[a,b]\times\Omega}|\uop_{2\sigma ml}(k\rv)|^2\diffV\\		
	=\frac{1}{k^2}\Re\left[\hop^{(2)}_l(kr)\Rop_{2l}^{(3)}(kr)r\right]_a^{b}
	+\int_{a}^{b}|\hop^{(1)}_l(kr)|^2 r^2\diff r,
\end{equation}
where we have used the Wronskian relation $z^{\ast}z^{\prime}
	-z^{\prime\ast}z	=-2\iu/x^2$, for $z=\hop_l^{(1)}(x)$, and the recursion relations for the spherical Hankel functions~\cite{vanBladel2007} in the last steps. 
The terms can be evaluated as $b\rightarrow\infty$ by considering the asymptotic behavior of the spherical Hankel functions.

\section{Q from Brune synthesis}\label{S:Brune}
Q factors can be determined from the stored energy and dissipated power in circuit models of the antenna. Chu used circuit models of the spherical modes to determine physical bounds on antennas~\cite{Chu1948}. The models can be adapted to small antennas that express the Q in the lumped circuit elements~\cite{Gustafsson2010b}. For general antennas, we synthesize an equivalent circuit from the input impedance using Brune synthesis~\cite{Brune1931}. There are several approaches to synthesize circuit models, \eg Brune, Bott and Duffin, Miyata, or Darlington synthesis~\cite{Wing2008}. The synthesis methods can produce different circuit so the circuits are not unique. It is also possible to synthesize lumped circuits with an internal stored energy that is non-observable from the input impedance. 

The circuit synthesis is based on expressing the input impedance as a (rational) positive real (PR) function in the complex frequency variable $s=\sigma+\ju\omega$ and subsequent manipulation of the PR function to identify the circuit elements. This requires modeling over wide bandwidths and shows that the resulting Q depends on the global properties of the input impedance.  

We start to construct a rational approximation of the antenna input impedance. 
In the range $\omega_1\leq\omega\leq\omega_2$ we use a rational function of order $(m_1,m_2)$, with $|m_1-m_2|\leq 1$, that is fitted to the input impedance using the MATLAB function \verb+invfreqs+. The order is chosen as low as possible such that the relative error is below some threshold, here we use $10^{-3}$, and that the rational function is a PR function~\cite{Wing2008}. 
\begin{figure}[t]
\begin{center}
  \includegraphics[width=0.48\textwidth]{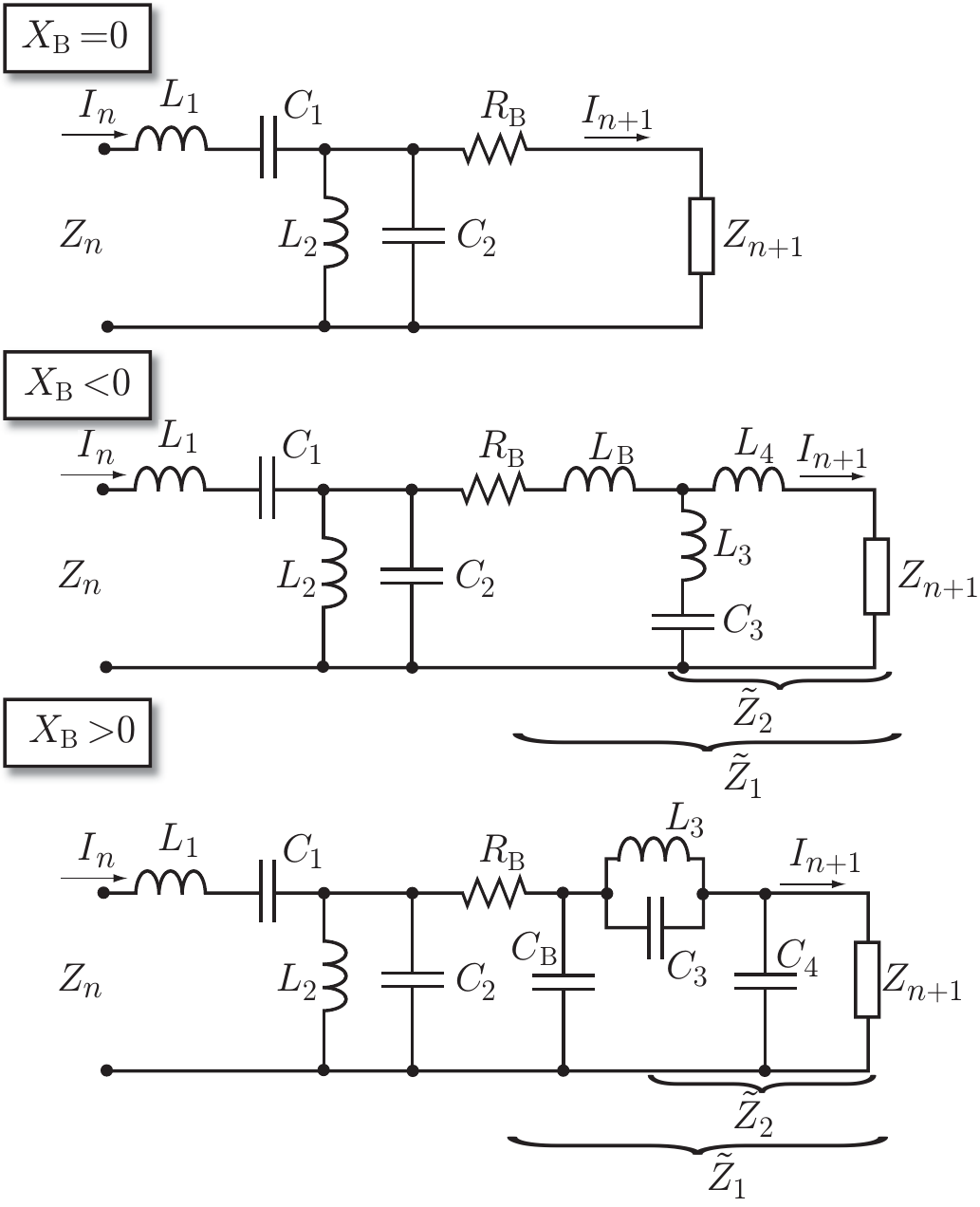}
\end{center}	
  \caption{Illustration of the Brune synthesis of a lumped circuit from a rational positive real (PR) input impedance $Z_{\mrm{in}}=Z_1$. In each iteration step, $Z_{n}\to Z_{n+1}$, series and shunt capacitance and inductance are first removed. Then the series resistance $R_{\mrm{B}}=\min_{\omega}\Re Z$ is removed. This leaves a PR function with $Z(\omega_0)=\ju X_{\mrm{B}}$. Depending on the sign of $X_{\mrm{B}}=\Im Z(\omega_0)$ a negative inductance or capacitance is removed. Finally, a resonance LC circuit and series inductance or shunt capacitance are removed, see~\cite{Brune1931,Wing2008} for details. Note, that $R_{\mrm{B}}=0$ and $X_{\mrm{B}}=0$ are treated separately.}
  \label{fig:BruneSynthesis}
\end{figure}

Brune synthesis~\cite{Brune1931,Wing2008} is an iterative procedure, where the order of the rational PR function modeling the input impedance is reduced in each step $Z_n\to Z_{n+1}$, see Fig.~\ref{fig:BruneSynthesis}. Here, we only present a brief overview of the Brune synthesis for the purpose of calculating the stored energy, see~\cite{Brune1931,Wing2008} for details.
First, series ($C_1,L_1$) and shunt ($C_2,L_2$) capacitance and inductance are removed by identification of the asymptotic expansion of the input impedance and admittance at $s=0$ and $s=\infty$. Then, the series resistance $R_{\mrm{B}}=\min_{s=\ju\omega}\Re Z$ is removed. This leaves a PR function with $Z(\ju\omega_0)=\ju X_{\mrm{B}}$. Depending on the sign of $X_{\mrm{B}}=\Im Z(\ju\omega_0)$ a negative inductance or capacitance is removed, see Fig.~\ref{fig:BruneSynthesis}. Finally, a resonance LC circuit and series inductance or shunt capacitance are removed, see~\cite{Brune1931,Wing2008} for details. Note, that also $R_{\mrm{B}}=0$ and $X_{\mrm{B}}=0$ are treated separately. This leaves a PR function, $Z_{n+1}$ of lower order than $Z_n$. The iteration, $Z_n\to Z_{n+1}$, is terminated when a pure resistive load remains, \ie $\Im Z_{n+1}=0$.

\begin{figure}[t]
\begin{center}
  \includegraphics[width=0.4\textwidth]{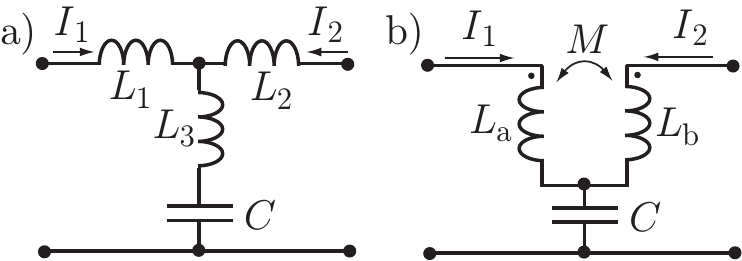}
\end{center}	
  \caption{Transformation of the \textit{T} network (a) to an ideal transformer (b).}
  \label{fig:tee2transf}
\end{figure}
The stored energy is easily calculated in the iterative synthesis procedure. The stored electric and magnetic energy in a capacitor, $C$, and inductor, $L$, are $\WE = |V|^2 C/4$ and $\WM = |I|^2 L/4$, respectively, where $I$ and $V$ denote the current and voltage, respectively. 
For simplicity, consider the case with a series inductor, \ie $Z_{\mrm{B}}=\ju\omega L_{\mrm{B}}$.
The stored electric and magnetic energies are then iteratively given by
\begin{equation}
	\We_{\mrm{B},n} = \frac{|I_n|^2}{4\omega C_1} + \frac{|\widetilde{V}_1|^2 C_2}{4}
	+\frac{|\widetilde{V}_2|^2}{4\omega C_3(\omega L_3-\frac{1}{\omega C_3})^2}
	+\We_{\mrm{B},n+1}
\label{eq:WeB}
\end{equation}
and
\begin{equation}
	\Wm_{\mrm{B},n} = \frac{L_1 |I_n|^2}{4} + \frac{|\widetilde{V}_1|^2}{4\omega L_2}
	+\frac{|\widetilde{V}_1|^2L_{\mrm{B}}}{4|\widetilde{Z}|_1^2}
	+\frac{|\widetilde{V}_2|^2L_{3}}{4(\omega L_3-\frac{1}{\omega C_3})^2}
	+\frac{|I_{n+1}|^2L_4}{4}
	+\Wm_{\mrm{B},n+1},
\label{eq:WmB}
\end{equation}
where $\widetilde{V}_n$ is the voltage over $\widetilde{Z}_n$ for $n=1,2$. One problem with the Brune synthesis is that it uses negative inductors and capacitors~\cite{Brune1931,Wing2008}, see $L_{\mrm{B}}$ and $C_{\mrm{B}}$ in Fig.~\ref{fig:BruneSynthesis}. This is resolved by transforming the \textit{T} containing the negative element to an ideal transformer.

The transformation of a \textit{T} network with arbitrary inductors $L_m$, $m=1,2,3$ and a capacitor $C$ to an ideal transformer is illustrated in Fig.~\ref{fig:tee2transf}. We assume that $L_1<0$ is negative corresponding to the Brune inductance $L_{\mrm{B}}$ in Fig.~\ref{fig:BruneSynthesis}. The inductors in the ideal transformer are $M=L_3$, $L_{\mrm{a}}=L_1+L_3$, and $L_{\mrm{b}}=L_2+L_3$. From the Brune synthesis~\cite{Brune1931,Wing2008}, the inductors are related as $L_1L_2+L_1L_3+L_2L_3=0$, and hence $L_{\mrm{a}}L_{\mrm{b}}=L_3^2$ showing that $L_{\mrm{a}}>0$. The stored magnetic energy in the \textit{T} and ideal transformer in Fig.~\ref{fig:tee2transf} is 
\begin{equation}
	\Wm = \frac{|I_1|^2 L_1}{4} + \frac{|I_2|^2 L_2}{4} + \frac{|I_1+I_2|^2 L_3}{4}
	=\frac{|I_1|^2 L_{\mrm{a}}}{4} + \frac{|I_2|^2 L_{\mrm{b}}}{4} + \frac{\Re\{I_1 I_2^{\ast}\}M}{2}.
\label{eq:WmB1}
\end{equation}
The corresponding stored electric energy is $\We=|I_1+I_2|^2/(4\omega^2 C)$ showing that the stored energy in the ideal transformer is identical to the stored energy in the original circuit representation. 

\begin{figure}[t]
\begin{center}
  \includegraphics[width=0.4\textwidth]{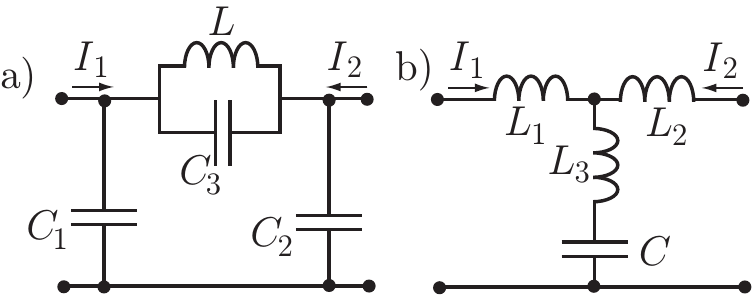}
\end{center}	
  \caption{Transformation of the $\Pi$ network (a) to a \textit{T} network (b).}
  \label{fig:Pi2tee}
\end{figure}
The interpretation for the case $X_{\mrm{B}}>0$ is more involved~\cite{Brune1931,Wing2008}. Here, the $\Pi$ network with the capacitors $C_1,C_2,C_3$ are first transform to a \textit{T} network with the components $C=C_1+C_2$, $L_1=LC_2/C$, $L_2=LC_1/C$, $L_3=LC_3/C$, see Fig.~\ref{fig:Pi2tee}. The stored energies are not the same in the $\Pi$ and \textit{T} networks for general component values. However, the components are not arbitrary in the Brune synthesis~\cite{Brune1931,Wing2008}. The capacitors in the $\Pi$ network are constrained as $C_1C_2+C_1C_3+C_2C_3=0$ and the inductors in the  \textit{T} network satisfy $L_1L_2+L_1L_3+L_2L_3=0$. The stored energies are identical for this case. The \textit{T} network is finally transform to an ideal transformer, see Fig.~\ref{fig:tee2transf}. We consider the stored electric energy to see that the stored energies are identical for the $\Pi$ and \textit{T} networks in the Brune case, see Fig.~\ref{fig:Pi2tee}. With prescribed voltages $V_1$ and $V_2$, we have 
\begin{multline}
	4\We = |V_1|^2C_1+|V_2|^2C_2+|V_1-V_2|^2C_3
	=|V_1|^2C_1+|V_2|^2C_2-|V_1-V_2|^2\frac{C_1C_2}{C_1+C_2}\\
	=\frac{|C_1V_1+C_2V_2|^2}{C_1+C_2}
	=\frac{|C_1V_1+C_2V_2|^2}{C}
\label{eq:WeBrune1}
\end{multline}
for the $\Pi$ network, where $C_1C_2+C_1C_3+C_2C_3=0$ is used. The stored energy in the corresponding \textit{T} network is
\begin{equation}
	4\We = \frac{|V|^2}{|\omega L_3-\frac{1}{\omega C}|^2 \omega^2 C}
	=\frac{|V|^2C}{|\omega^2 LC_3-1|^2}
\label{eq:WeBrune2}
\end{equation}
where $V$ is the voltage
\begin{equation}
	V = \frac{\frac{V_1}{\ju\omega L_1}+\frac{V_2}{\ju\omega L_2}}{\frac{1}{\ju\omega L_1}+\frac{1}{\ju\omega L_2}+\frac{1}{\ju\omega L_3+\frac{1}{\ju\omega C}}}
	=\frac{C_1V_1+C_2V_2}{C+\frac{C_1C_2}{C_3-\frac{1}{\omega^2 L}}}
	=(C_1V_1+C_2V_2)\frac{\omega^2LC_3-1}{C}
\label{eq:VBrune}
\end{equation}
that inserted into~\eqref{eq:WeBrune2} gives~\eqref{eq:WeBrune1} and shows that the stored electric energies are identical for the $\Pi$ and \textit{T} networks for the case $C_1C_2+C_1C_3+C_2C_3=0$. The \textit{T} network has a negative inductance $L_2$ that is removed in the final transformation to the ideal transformer, see Fig.~\ref{fig:tee2transf}. 

\section{Bode-Fano matching limitations}\label{S:BodeFano}
We consider the Bode-Fano matching to determine a bound for lossless matching networks, see~\cite{Fano1950} for details. Using the asymptotic expansions and the zeros in the complex half plane  
for the shunt LC series LC circuit in Fig.~\ref{fig:BQ4pLCsLCR}, we have the integral identities 
\begin{align}
	\frac{2}{\pi}\int_0^{\infty}\omega^2\ln\frac{1}{|\refl(\omega)|}\diff\omega &= \frac{2}{Q_{\mrm{c}}} +\frac{2}{3}\Re\sum_{n} \lambda_n^3\\
	\frac{2}{\pi}\int_0^{\infty}\ln\frac{1}{|\refl(\omega)|}\diff\omega &= \frac{2}{Q_{\mrm{g}}} -2\Re\sum_{n} \lambda_n\\
	\frac{2}{\pi}\int_0^{\infty}\frac{1}{\omega^2}\ln\frac{1}{|\refl(\omega)|}\diff\omega &= \frac{2}{Q_{\mrm{g}}} -2\Re\sum_{n} \frac{1}{\lambda_n}\\
	\frac{2}{\pi}\int_0^{\infty}\frac{1}{\omega^4}\ln\frac{1}{|\refl(\omega)|}\diff\omega &= \frac{2}{Q_{\mrm{c}}} +\frac{2}{3}\Re\sum_{n} \frac{1}{\lambda_n^3}
\end{align}
where $Q_{\mrm{g}}=\max\{Q_{\mrm{p}},Q_{\mrm{s}}\}$, $Q_{\mrm{l}}=\min\{Q_{\mrm{p}},Q_{\mrm{s}}\}$, $Q_{\mrm{c}}=3Q_{\mrm{g}}^3 Q_{\mrm{l}}/(3Q_{\mrm{g}}^2Q_{\mrm{l}}+3Q_{\mrm{g}}-Q_{\mrm{l}})\leq Q_{\mrm{g}}$, $\Re\lambda_n\geq 0$, and we have assumed that $\omega_0=1$. 
We bound the integrals using $\max_{\omega}|\refl(\omega)|=\refl_0$ for $\omega\in\omega_0[1-B/2,1+B/2]$ giving the inequalities
\begin{align}
	\frac{1}{\pi}(B+B^3/12)\ln\frac{1}{|\refl_0|} &\leq \frac{1}{Q_{\mrm{c}}} +\frac{1}{3}\Re\sum_{n} \lambda_n^3\\
	\frac{1}{\pi}B\ln\frac{1}{|\refl_0|} &\leq \frac{1}{Q_{\mrm{g}}} -\Re\sum_{n} \lambda_n\\
	\frac{1}{\pi}\frac{B}{1-B^2/4}\ln\frac{1}{|\refl_0|}  &\leq \frac{1}{Q_{\mrm{g}}} -\Re\sum_{n} \frac{1}{\lambda_n}\\
	\frac{1}{\pi}\frac{B+B^3/12}{(1-B^2/4)^3}\ln\frac{1}{|\refl_0|} &\leq \frac{1}{Q_{\mrm{c}}} +\frac{1}{3}\Re\sum_{n} \frac{1}{\lambda_n^3}
\end{align}
We note that the middle equations are identical to the Bode-Fano bound for the RCL circuit~\cite{Gustafsson+Nordebo2006b} for maximal Q value $\max\{Q_{\mrm{p}},Q_{\mrm{s}}\}$. This is natural as the cascaded shunt (or series) circuit cannot improve the matching. It is also seen that a complex conjugate pair gives the optimal $\lambda_n$ for $B\ll 1$ and that this case reduces to the bound for the RCL circuit. The set of inequalities are solved numerically for $\refl_0$ given $B$ and assuming a complex conjugate pair $\lambda_n$.

\bibliographystyle{teorel}\bibliography{total,matsadd}

\begin{thebibliography}{10}

\bibitem{Bostrom+Kristensson+Strom1991}
A.~Bostr{\"o}m, G.~Kristensson, and S.~Str{\"o}m.
\newblock Transformation properties of plane, spherical and cylindrical scalar
  and vector wave functions.
\newblock In V.~V. Varadan, A.~Lakhtakia, and V.~K. Varadan, editors, {\em
  Field Representations and Introduction to Scattering}, Acoustic,
  Electromagnetic and Elastic Wave Scattering, chapter~4, pages 165--210.
  Elsevier Science Publishers, Amsterdam, 1991.

\bibitem{Brune1931}
O.~Brune.
\newblock Synthesis of a finite two-terminal network whose driving-point
  impedance is a prescribed function of frequency.
\newblock {\em MIT J. Math. Phys.}, {\bf 10}, 191--236, 1931.

\bibitem{Capek+etal2012}
M.~Capek, P.~Hazdra, and J.~Eichler.
\newblock A method for the evaluation of radiation {Q} based on modal approach.
\newblock {\em IEEE Trans. Antennas Propagat.}, {\bf 60}(10), 4556--4567, 2012.

\bibitem{Capek+etal2014}
M.~Capek, L.~Jelinek, P.~Hazdra, and J.~Eichler.
\newblock The measurable {Q} factor and observable energies of radiating
  structures.
\newblock {\em arXiv preprint arXiv:1309.6122}, 2013.

\bibitem{Carlin+Civalleri1998}
H.~J. Carlin and P.~P. Civalleri.
\newblock {\em Wideband circuit design}.
\newblock CRC Press, Boca Raton, 1998.

\bibitem{Carpenter1989}
C.~J. Carpenter.
\newblock Electromagnetic energy and power in terms of charges and potentials
  instead of fields.
\newblock {\em IEE Proc.~A}, {\bf 136}(2), 55--65, 1989.

\bibitem{Chu1948}
L.~J. Chu.
\newblock Physical limitations of omni-directional antennas.
\newblock {\em J. Appl. Phys.}, {\bf 19}, 1163--1175, 1948.

\bibitem{TEAT-7227}
M.~Cismasu and M.~Gustafsson.
\newblock Antenna bandwidth optimization with single frequency simulation.
\newblock Technical Report LUTEDX/(TEAT-7227)/1--28/(2013), Lund University,
  Department of Electrical and Information Technology, P.O. Box 118, S-221 00
  Lund, Sweden, 2013.
\newblock http://www.eit.lth.se.

\bibitem{Collin1998}
R.~E. Collin.
\newblock Minimum {Q} of small antennas.
\newblock {\em J. Electro. Waves Applic.}, {\bf 12}, 1369--1393, 1998.

\bibitem{Collin+Rothschild1964}
R.~E. Collin and S.~Rothschild.
\newblock Evaluation of antenna {Q}.
\newblock {\em IEEE Trans. Antennas Propagat.}, {\bf 12}, 23--27, January 1964.

\bibitem{Endean+Carpenter1992}
V.~G. Endean and C.~J. Carpenter.
\newblock Electromagnetic energy and power in terms of charges and potentials
  instead of fields (comments with reply).
\newblock In {\em IEE Proc.~A}, volume 139, pages 338--342. IET, 1992.

\bibitem{Fano1950}
R.~M. Fano.
\newblock Theoretical limitations on the broadband matching of arbitrary
  impedances.
\newblock {\em Journal of the Franklin Institute}, {\bf 249}(1,2), 57--83 and
  139--154, 1950.

\bibitem{Feynman1965}
R.~P. Feynman, R.~B. Leighton, and M.~Sands.
\newblock {\em The Feynman Lectures on Physics}.
\newblock Addison-Wesley, Reading, MA, USA, 1965.

\bibitem{Foltz+McLean1999}
H.~D. Foltz and J.~S. McLean.
\newblock Limits on the radiation {Q} of electrically small antennas restricted
  to oblong bounding regions.
\newblock In {\em IEEE Antennas and Propagation Society International
  Symposium}, volume~4, pages 2702--2705. IEEE, 1999.

\bibitem{Geyi2003}
W.~Geyi.
\newblock Physical limitations of antenna.
\newblock {\em IEEE Trans. Antennas Propagat.}, {\bf 51}(8), 2116--2123, August
  2003.

\bibitem{Gustafsson2010b}
M.~Gustafsson.
\newblock Sum rules for lossless antennas.
\newblock {\em IET Microwaves, Antennas \& Propagation}, {\bf 4}(4), 501--511,
  2010.

\bibitem{Gustafsson+Nordebo2013}
M.~Gustafsson and S.~Nordebo.
\newblock Optimal antenna currents for {Q}, superdirectivity, and radiation
  patterns using convex optimization.
\newblock {\em IEEE Trans. Antennas Propagat.}, {\bf 61}(3), 1109--1118, 2013.

\bibitem{Gustafsson+etal2012a}
M.~Gustafsson, M.~Cismasu, and B.~L.~G. Jonsson.
\newblock {Physical bounds and optimal currents on antennas}.
\newblock {\em IEEE Trans. Antennas Propagat.}, {\bf 60}(6), 2672--2681, 2012.

\bibitem{Gustafsson+Nordebo2006b}
M.~Gustafsson and S.~Nordebo.
\newblock Bandwidth, {Q} factor, and resonance models of antennas.
\newblock {\em Progress in Electromagnetics Research}, {\bf 62}, 1--20, 2006.

\bibitem{Hansen1988}
J.~E. Hansen, editor.
\newblock {\em Spherical Near-Field Antenna Measurements}.
\newblock Number~26 in {IEE} electromagnetic waves series. Peter Peregrinus
  Ltd., Stevenage, UK, 1988.
\newblock {ISBN}: 0-86341-110-X.

\bibitem{Hansen+Collin2009}
R.~C. Hansen and R.~E. Collin.
\newblock A new {C}hu formula for {Q}.
\newblock {\em IEEE Antennas and Propagation Magazine}, {\bf 51}(5), 38--41,
  2009.

\bibitem{Hansen+etal2012}
T.~V. Hansen, O.~S. Kim, and O.~Breinbjerg.
\newblock Stored energy and quality factor of spherical wave functions--in
  relation to spherical antennas with material cores.
\newblock {\em IEEE Trans. Antennas Propagat.}, {\bf 60}(3), 1281--1290, 2012.

\bibitem{Harrington1961}
R.~F. Harrington.
\newblock {\em Time Harmonic Electromagnetic Fields}.
\newblock McGraw-Hill, New York, 1961.

\bibitem{Hazdra+etal2011}
P.~Hazdra, M.~Capek, and J.~Eichler.
\newblock Radiation {Q}-factors of thin-wire dipole arrangements.
\newblock {\em Antennas and Wireless Propagation Letters, IEEE}, {\bf 10},
  556--560, 2011.

\bibitem{Jackson1999}
J.~D. Jackson.
\newblock {\em Classical Electrodynamics}.
\newblock John Wiley \& Sons, New York, third edition, 1999.

\bibitem{Jin2010}
J.~M. Jin.
\newblock {\em Theory and computation of electromagnetic fields}.
\newblock Wiley Online Library, 2010.

\bibitem{Landau+Lifshitz1984}
L.~D. Landau, E.~M. Lifshitz, and L.~P. Pitaevski\u{\i}.
\newblock {\em Electrodynamics of Continuous Media}.
\newblock Pergamon, Oxford, second edition, 1984.

\bibitem{McLean1996}
J.~S. McLean.
\newblock A re-examination of the fundamental limits on the radiation {$Q$} of
  electrically small antennas.
\newblock {\em IEEE Trans. Antennas Propagat.}, {\bf 44}(5), 672--676, May
  1996.

\bibitem{Rahmat-Samii+Michielssen1999}
Y.~Rahmat-Samii and E.~Michielssen.
\newblock {\em Electromagnetic Optimization by Genetic Algorithms}.
\newblock Wiley Series in Microwave and Optical Engineering. John Wiley \&
  Sons, 1999.

\bibitem{Sten+etal2001}
J.~C.-E. Sten, P.~K. Koivisto, and A.~Hujanen.
\newblock Limitations for the radiation {$Q$} of a small antenna enclosed in a
  spheroidal volume{:} axial polarisation.
\newblock {\em AE\"U Int. J. Electron. Commun.}, {\bf 55}(3), 198--204, 2001.

\bibitem{Stuart+etal2007}
H.~Stuart, S.~Best, and A.~Yaghjian.
\newblock Limitations in relating quality factor to bandwidth in a double
  resonance small antenna.
\newblock {\em Antennas and Wireless Propagation Letters}, {\bf 6}, 2007.

\bibitem{Thal1978}
H.~L. Thal.
\newblock Exact circuit analysis of spherical waves.
\newblock {\em IEEE Trans. Antennas Propagat.}, {\bf 26}(2), 282--287, March
  1978.

\bibitem{Thal2006}
H.~L. Thal.
\newblock New radiation {Q} limits for spherical wire antennas.
\newblock {\em IEEE Trans. Antennas Propagat.}, {\bf 54}(10), 2757--2763,
  October 2006.

\bibitem{Uehara+etal1992}
M.~Uehara, J.~E. Allen, and C.~J. Carpenter.
\newblock Electromagnetic energy and power in terms of charges and potentials
  instead of fields (comments with reply).
\newblock In {\em IEE Proc.~A}, volume 139, pages 42--44. IET, 1992.

\bibitem{vanBladel2007}
J.~G. Van~Bladel.
\newblock {\em Electromagnetic Fields}.
\newblock IEEE Press, Piscataway, NJ, second edition, 2007.

\bibitem{Vandenbosch2010}
G.~A.~E. Vandenbosch.
\newblock Reactive energies, impedance, and {Q} factor of radiating structures.
\newblock {\em IEEE Trans. Antennas Propagat.}, {\bf 58}(4), 1112--1127, 2010.

\bibitem{Vandenbosch2011}
G.~A.~E. Vandenbosch.
\newblock Simple procedure to derive lower bounds for radiation {Q} of
  electrically small devices of arbitrary topology.
\newblock {\em IEEE Trans. Antennas Propagat.}, {\bf 59}(6), 2217--2225, 2011.

\bibitem{Vandenbosch2013a}
G.~A.~E. Vandenbosch.
\newblock Radiators in time domain, part {I}: electric, magnetic, and radiated
  energies.
\newblock {\em IEEE Trans. Antennas Propagat.}, {\bf 61}(8), 3995--4003, 2013.

\bibitem{Vandenbosch2013b}
G.~A.~E. Vandenbosch.
\newblock Radiators in time domain, part {II}: finite pulses, sinusoidal regime
  and {Q} factor.
\newblock {\em IEEE Trans. Antennas Propagat.}, {\bf 61}(8), 4004--4012, 2013.

\bibitem{Volakis+etal2010}
J.~Volakis, C.~C. Chen, and K.~Fujimoto.
\newblock {\em Small Antennas: Miniaturization Techniques \& Applications}.
\newblock McGraw-Hill, New York, 2010.

\bibitem{Wing2008}
O.~Wing.
\newblock {\em Classical Circuit Theory}.
\newblock Springer, New York, 2008.

\bibitem{Yaghjian+Best2005}
A.~D. Yaghjian and S.~R. Best.
\newblock Impedance, bandwidth, and {$Q$} of antennas.
\newblock {\em IEEE Trans. Antennas Propagat.}, {\bf 53}(4), 1298--1324, 2005.

\end{thebibliography}


\end{document}